%% file: main.tex
\documentclass[%
reprint,
 amsmath,amssymb,
 aps,
 prl,
]{revtex4-1}

\usepackage{xcolor}
\usepackage{graphicx}
\usepackage{dcolumn}
\usepackage{bm}
\usepackage{hyperref}

\usepackage[normalem]{ulem} 

\begin{document}

\title{Non-reciprocal alignment induces asymmetric clustering in active mixtures}

\author{Kim L. Kreienkamp}
\email{k.kreienkamp@tu-berlin.de}
\author{Sabine H. L. Klapp}
\email{sabine.klapp@tu-berlin.de}
\affiliation{%
 Institut f\"ur Theoretische Physik, Technische Universit\"at Berlin
}%


\begin{abstract}
Heterogeneity is ubiquitous in biological and synthetic active matter systems that are inherently out of equilibrium. Typically, such active mixtures involve not only conservative interactions between the constituents, but also non-reciprocal couplings, whose full consequences for the collective behavior still remain elusive.
Here, we study a minimal active non-reciprocal mixture with both, symmetric isotropic and non-reciprocal polar interactions. By combining a hydrodynamic theory derived from microscopic equations and particle-based simulations, we provide a scale-bridging view on the rich dynamics that occur even in absence of oscillatory instabilities. We show, in particular, that non-reciprocal alignment alone induces {\it asymmetrical} clustering at otherwise fully symmetric parameters. These density inhomogeneities go beyond the typical band formation in Vicsek-like systems. Within the asymmetric clustering state, single-species clusters chase more dilute accumulations of the other species.
\end{abstract}

\maketitle


The phase behavior of fluid mixtures and, in particular, their spontaneous demixing,
has been fascinating researchers for decades \cite{rowlinson_2013_liquids_mixture_book}. In thermal equilibrium, demixing is ruled by energy and entropy. It occurs, e.g., when particles differ in shape \cite{adams_fraden_1998_entropically_driven_phase_transitions_mixtures_rods_spheres} or size \cite{asakura_oosawa_1958_suspended_particles_in_macromolecules,biben_frenkel_1996_depletion_effects}, or if interactions between different species are weak against attraction within the same species \cite{wilding_nielaba_1998_liquid-vapor_phase_behavior_symmetrical_binary_fluid_mixture}. The situation
becomes more complex in living and active systems that are inherently out of equilibrium and often heterogeneous, examples being bacterial colonies and swarms \cite{elias_banin_2012_multispecies_biofilms,ben-jacob_ingham_2016_multispecies_swarms_social_microorganisms,peled_beer_2021_heterogeneous_bacterial_swarms}, synthetic active-passive mixtures \cite{leptos_goldstein_2009_tracer_diffusion_in_bacteria,stenhammar_cates_2015_mips_active_passive_mixture}, and membranes \cite{zhao_zhang_2020_phase_separation_membrane,sens_turner_2011_microphase_separation}. 

Already one-component active systems exhibit intriguing out-of-equilibrium states like flocking 
\cite{marchetti_simha_2013_hydrodynamics_soft_active_matter,vicsek_1995_novel_type_phase_transition,gregoire_chate_2004_collective_motion}, motility-induced phase separation (MIPS) \cite{cates_tailleur_2015_mips,Bialke_2013_microscopic_theory_phase_seperation}, and clustering \cite{buttinoni_2013_dynamical_clustering}. This indicates an even richer dynamics of active {\it mixtures}.
Yet, although certain aspects of their self-organization have been studied \cite{Adhikary_Santra_2022_pattern_formation_phase_transition_binary_mixture,kolb_2020_active_binary_mixture_slow_fast,ilker_joanny_2020_phase_separation_mixtures_different_temperatures,weber_frey_2016_binary_mixture_different_diffusivities_demix,peled_beer_2021_heterogeneous_bacterial_swarms,maity_morin_2023_spontaneous_demixing_binary_colloidal_flocks}, 
we are still far away from a comprehensive understanding.

Active mixtures often contain non-reciprocally coupled particles, whose interactions apparently break action-reaction symmetry. Non-reciprocal couplings are a common feature in various areas of nonequilibrium physics, including predator-prey systems \cite{lotka_1920_lotka_volterra_model,volterra_1926_lotka_volterra_model,meredith_zarzar_2020_predator-prey_droplets}, neural networks \cite{sompolinsky_kanter_1986_temporal_association_asymmetric_neural_networks,rieger_zittartz_1989_glauber_dynamics_asymmetric_SK_model}, systems with vision cones \cite{barberis_peruani_2016_minimal_cognitive_flocking_model,lavergne_bechinger_2019_group_formation_visual_perception-dependent_motility,loos_martynec_2023_long-range_order_non-reciprocal_XY_model,avni_vitelli_2023_non-reciprocal_Ising_model}, odd solids \cite{scheibner_vitelli_2020_odd_elasticity,fruchart_vitelli_2023_odd_viscosity_and_elasticity}, and quantum optics \cite{metelmann_2015_nonreciprocal_photon_transmission,zhang_2018_non-reciprocal_quantum_optical_system,mcdonald_2022_nonequilibrium_quantum_non-Hermitian_lattice}. On the particle scale, non-reciprocal interactions emerge from nonequilibrium environments \cite{scheibner_vitelli_2020_odd_elasticity,gupta_ramaswamy_2022_active_nonreciprocal_attraction_elastic_medium},	e.g., by phoresis \cite{agudo-canalejo_golestanian_2019_active_phase_separation_chemically_interacting_particles,soto_2014_self-assembly_catalytically_active_colloidal_molecules,saha_golestanian_2019_pairing_waltzing_scattering_chemotactic_active_colloids,tucci_saha_2024_nonreciprocal_collective_dynamics_mixtures_Janus}, fluid flow \cite{maity_morin_2023_spontaneous_demixing_binary_colloidal_flocks,banerjee_rao_2022_nonreciprocity_compressible_viscoelastic_fluid,gupta_ramaswamy_2022_active_nonreciprocal_attraction_elastic_medium}, or quorum sensing \cite{duan_mahault_2023_dynamical_pattern_formation_without_self-attraction,knevzevic_stark_2022_collective_motion_non-reciprocal_orientational_interactions,vicsek_1995_novel_type_phase_transition}.
Recent field-theoretical studies \cite{You_Baskaran_Marchetti_2020_pnas,fruchart_2021_non-reciprocal_phase_transitions,saha_scalar_active_mixtures_2020,frohoff_thiele_2023_nonreciprocal_cahn-hilliard,Frohoff-Huelsmann_Thiele_2021_cahn-hilliard_nonvariational_coupling,hanai_2024_nonreciprocal_frustration} and particle-based simulations \cite{mandal_sollich_2022_robustness_travelling_states_generic_non-reciprocal_mixture,duan_mahault_2023_dynamical_pattern_formation_without_self-attraction} of simple mixtures with either scalar or polar order parameters 
have shown that non-reciprocal couplings can have drastic effects, including the spontaneous formation of time-dependent 
states \cite{You_Baskaran_Marchetti_2020_pnas,fruchart_2021_non-reciprocal_phase_transitions,saha_scalar_active_mixtures_2020}. However, typical active mixtures involve not only one, but several types of interactions, each potentially responsible for specific collective behaviors (e.g., flocking or MIPS) on its own. This yields a new class of non-reciprocal systems with mixed order parameters and, thus, potentially unknown emerging dynamics. Recent experiments indicate unexpected phase separation in non-reciprocal mixtures of polar Quincke rollers \cite{maity_morin_2023_spontaneous_demixing_binary_colloidal_flocks}, and in mixtures of repulsive robots \cite{chen_zhang_2024_emergent_chirality_hyperuniformity_active_mixture_NR}.
Inspired by these findings we here ask the question: How do non-reciprocal polar couplings {\it alone} affect systems featuring clustering and MIPS? Can they even induce demixing?

To this end we consider a minimal model of a binary mixture of circular active Brownian particles (ABPs). The non-reciprocal character enters only via the interspecies alignment which may be asymmetric and even opposing. Apart from that, repulsive interactions are symmetric (favoring MIPS) and intraspecies alignment is reciprocal (favoring flocking within each species). 
Without repulsion, our model reduces to a polar non-reciprocal mixture of point-like particles that, at large coupling strengths, exhibits parity-time-symmetry breaking \cite{fruchart_2021_non-reciprocal_phase_transitions}. 
We here explore the density dynamics in the regime below the threshold of parity-time-symmetry-breaking. We find, that non-reciprocal alignment leads to {\it asymmetric} density dynamics manifested by the formation of clusters of only {\it one} species, akin to partial demixing.
These spatial inhomogeneities are strongly different from those arising from pure polar couplings, such as high-density polarized bands \cite{vicsek_1995_novel_type_phase_transition,gregoire_chate_2004_collective_motion,solon_tailleur_2015_phase_separation_flocking_models,chatterjee_noh_2023_flocking_unfriendly_species}. Here, we provide a scale-bridging view on the preferential clustering of only one species due to non-reciprocal alignment. Based on a mean-field continuum theory, we characterize the rich dynamics of our active mixture by nonequilibrium phase diagrams. In addition, we perform particle-based simulations and a corresponding fluctuation analysis to unravel effects beyond mean-field theory. In particular, we find that the single-species cluster dynamics is characterized by chase-and-run behavior.

\textit{Model.}---We consider two species of circular ABPs with densities $\rho_0^a$ ($a=A,B$). The particles with positions $\bm{r}^a_i$ and heading vectors $\bm{p}^a_{i}=(\cos\theta^a_i, \sin\theta^a_i)^{\rm T}$ interact via hard steric repulsion $\bm{F}_{\rm{rep}}$ \cite{weeks_1971_Weeks-Chandler-Andersen_potential} and Vicsek-like torques.
The overdamped Langevin equations governing the dynamics, are given by
\begin{subequations} \label{eq:Langevin_eq}
	\begin{align}
		\dot{\bm{r}}^a_{i} &= v_0\,\bm{p}^a_{i} + \mu_{r} \sum_{j,b} \bm{F}_{\rm{rep}}(\bm{r}^a_{i},\bm{r}^b_{j}) + \sqrt{2\,D'_{\rm t}}\,\bm{\xi}_{i}^a \label{eq:Langevin_r}\\
		\dot{\theta}^a_{i} &= \mu_{\theta} \sum_{j,b \in \Omega_i(R_{\theta})} k_{ab}\, \sin(\theta_{j}^b-\theta_{i}^a) + \sqrt{2\,D'_{\rm r}}\,\eta_{i}^a \label{eq:Langevin_theta}.
	\end{align}
\end{subequations}
The two species have equal self-propulsion velocities $v_0$, equal translational ($\bm{\xi}_i^a(t)$) and rotational ($\bm{\eta}_i^a(t)$) unit-variance Gaussian white noises with zero mean, equal mobilities ($\mu_r$, $\mu_{\theta}$), and fully symmetric repulsive interactions. The only difference between the two species lies in their torques of strength $k_{ab}$, which can be positive or negative. Particles of species $a$ tend to orient parallel (align) or antiparallel (antialign) with $b$-particles within radius $R_{\theta}$ when $k_{ab}>0$ or $k_{ab}<0$, respectively. \textit{Reciprocal} couplings are defined by the choice $k_{AB} = k_{BA}$. We specifically allow for \textit{non-reciprocal} orientational couplings, for which $k_{AB}\neq k_{BA}$. The particle diameter $\ell = \sigma$ and the time $\tau = \sigma^2/D'_{\rm t}$ a (passive) particle needs to travel over its own distance are taken as characteristic length and time scales. The control parameters are then the densities $\rho_0^a$, the reduced orientational coupling parameters $g_{ab} = k_{ab}\,\mu_{\theta}\,\tau$, the P\'eclet number $\rm{Pe}=v_0\,\tau/\ell$, and rotational noise strength $D_{\rm r}=D'_{\rm r}\,\tau$. The strength of the hard repulsive potential and the cut-off distance for the torque, $R_{\theta}=2\,\ell$, are set constant. For the detailed equations of motion and corresponding Brownian Dynamics (BD) simulations, see \cite{SM}.

To understand large-scale pattern formation, we coarse-grain the microscopic dynamics in a mean-field approximation \cite{kreienkamp_klapp_2022_clustering_flocking_chiral_active_particles_non-reciprocal_couplings,fruchart_2021_non-reciprocal_phase_transitions,teVrugt_Wittkowski_2023_derivation_predictive_field_theory} and incorporate a density-dependent velocity $v^{\rm eff}(\rho)$ \cite{Bialke_2013_microscopic_theory_phase_seperation,speck_2015_dynamical_mean_field_phase_separation}. We obtain a six-dimensional hydrodynamic description for the density fields $\rho^{a}(\bm{r},t)$ and polarization densities ${\bm w}^a(\bm{r},t)$ \cite{SM}. These equations, and the linear stability analysis of the disordered uniform phase $(\rho^a, \bm{w}^a) = (1, \bm{0})$ \footnote{Within the here considered regime of $g_{AA}=g_{BB}=3$, no further information is obtained by changing from the uniform disordered base state to the homogeneous flocking base state. A linear stability analysis around homogeneous flocking states with $(\rho^a, \bm{w}^a) = (1, \bm{w}_0^a)$, where $\bm{w}_0^A$ either parallel or anti-parallel to $\bm{w}_0^B$, yields the same long-wavelength ($k=0$) flocking instabilities but no information about clustering instabilities.} upon perturbations of wavenumber $k$, form the basis of our investigation. On the continuum level, the alignment strength scales with the average single-species density $\rho_0^b$ and enters as $g'_{ab} = g_{ab}\,R_{\theta}^2\,\rho_0^b/2$ \cite{SM}.

To concentrate on the effect of non-reciprocal interspecies torques on phase separation, we here focus on the case of equal densities $\rho_0^A=\rho_0^B=\rho_0/2$ (for more general parameter choices, see \cite{SM}). 
We choose the density ($\rho_0^a = 4/(5\,\pi)$), motility (${\rm Pe}=40$), and noise strength ($D_{\rm r} = 3\cdot 2^{-1/3}$) such that the system exhibits MIPS in the absence of any alignment couplings ($g_{ab}=0$ $\forall\, ab$). We then set the intraspecies alignment couplings equal, i.e., $g_{AA}=g_{BB}=g$, and vary $g_{AB}$, $g_{BA}$ independently.

\textit{Flocking behavior.}---The alignment couplings between particles can induce states with nonzero global polarization $P_{a}=\vert \bm{P}_{a} \vert =\vert N_{a}^{-1}\sum_{\alpha}^{N_a} \bm{p}_{\alpha} \vert > 0$. A flocking state is characterized by parallel orientations of $A$- and $B$-flocks, i.e., $\bm{P}_A || \bm{P}_B$. On the other hand, in an antiflocking state, the two species each form flocks, yet with antiparallel direction, i.e., $\bm{P}_A || -\bm{P}_B$. The emergence of polarized states is related to long-wavelength ($k=0$) fluctuations of polarizations. The density fields are conserved and thus their fluctuations vanish at $k=0$. The mean-field $k=0$-polarization dynamics are given by \cite{SM}
\begin{equation}
	\label{eq:mean-field_polarization_matrix_equation}
	\partial_t \begin{pmatrix}
		\bm{w}^A \\
		\bm{w}^B
	\end{pmatrix} = \begin{pmatrix}
	g'-D_{\rm r} - \frac{\bm{\mathcal{Q}}_A^2}{2 D_{\rm r}} & g'_{AB} \\
	g'_{BA} & g'-D_{\rm r} - \frac{\bm{\mathcal{Q}}_B^2}{2 D_{\rm r}}
\end{pmatrix} \cdot \begin{pmatrix}
		\bm{w}^A \\
		\bm{w}^B
	\end{pmatrix}
\end{equation}
with
\begin{equation}
	\bm{\mathcal{Q}}_a = g'\,\bm{w}^a + g'_{ab}\,\bm{w}^b \ \text{with} \ b \neq a.
\end{equation}
Representing the polarization fields in terms of amplitudes and phases, we find that the linear stability analysis yields a $2\times 2$-eigenvalue problem for the amplitudes with eigenvalues \cite{SM}
\begin{equation}
		\sigma^{\rm mf}_{1/2} = g' - D_{\rm r} \pm \sqrt{g'_{AB} \, g'_{BA}} .
\end{equation}
If ${\rm Re}(\sigma_{1/2}^{\rm mf})>0$, polarized states, i.e., flocking or antiflocking, occur. The corresponding eigenvector signals whether the system exhibits flocking or antiflocking.

\begin{figure}
	\includegraphics[width=\linewidth]{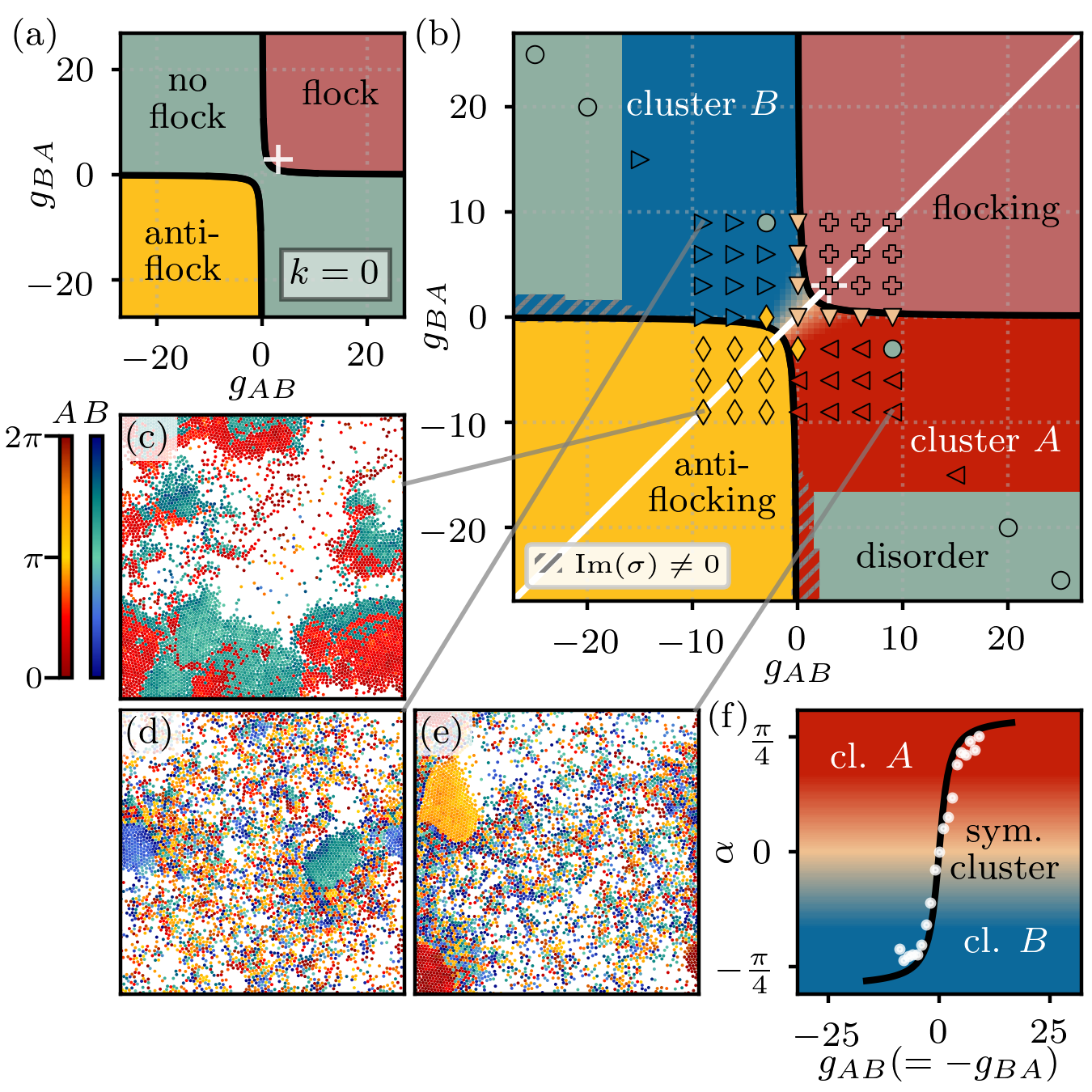}
	\caption{\label{fig:stability_diagram_with_snapshots}Nonequilibrium phase behavior at weak coupling ($g=3$). 
	(a) Stability diagram at $k=0$. (b) Full stability diagram from $6\times 6$-analysis (including $k>0$), revealing regions of (a)symmetric clustering. Color-coded marker points denote BD simulations. The white cross in (a,b) indicates the effective one-component system. (c-e): BD snapshots for (c) $g_{AB}=g_{BA}=-9$, (d) $g_{AB}=-g_{BA}=-9$, and (e) $g_{AB}=-g_{BA}=9$. Color code indicating particle type and orientation is provided in (c). (f) Clustering angle $\alpha$ from continuum predictions (line) and BD data (dots).}
\end{figure}

Henceforth, we will focus on a ``weak-intraspecies-coupling'' regime with $g' - D_{\rm r}<0$ by setting $g=3$. The corresponding mean-field phase diagram (at $k=0$) is shown in Fig.~\ref{fig:stability_diagram_with_snapshots}(a). The system only exhibits polarized states when the product $g'_{AB}\,g'_{BA}$ is large enough. For small $g'_{AB}\,g'_{BA}$, flocking and antiflocking are suppressed by rotational diffusion. (Anti-)flocking occurs if both $g'_{AB}$, $g'_{BA}$ are (negative) positive. These predictions are consistent with BD simulations [Fig.~\ref{fig:stability_diagram_with_snapshots}(c)].

\textit{Clustering behavior.}---At $k>0$, unstable density dynamics come into play \cite{kreienkamp_klapp_2022_clustering_flocking_chiral_active_particles_non-reciprocal_couplings}. To this end, we now go back to the full six-dimensional eigenvalue problem at arbitrary wavenumber $k$. The phase diagram is shown in Fig.~\ref{fig:stability_diagram_with_snapshots}(b). Within the here considered weak-intraspecies-coupling regime, particularly interesting density instabilities mainly occur outside the (anti-)flocking regime \cite{SM}.

It turns out that these density instabilities can be understood in terms of a simplified coarse-grained picture when considering large length and time scales in the non-flocking regime. This allows for adiabatic elimination of $\bm{w}^a$, yielding $\bm{w}_{\rm ad}^a= \bm{w}_{\rm ad}^a(\rho^A,\rho^B)$ \cite{SM}. The coarse-grained density dynamics then evolve as
\begin{equation}
	\partial_t \rho^a = - \nabla \cdot \big( v^{\rm eff}(\rho^A + \rho^B) \, \bm{w}_{\rm ad}^a \big) + D_{\rm t} \, \nabla^2 \rho^a .
\end{equation}
Linearizing around the homogeneous phase yields the eigenvalue equation \cite{SM}
\begin{equation}
	\sigma_{\rho} \begin{pmatrix}
		\hat{\rho}_A + \hat{\rho}_B \\
		\hat{\rho}_A - \hat{\rho}_B
	\end{pmatrix} = \bm{\mathcal{M}}_{\rm ad} \cdot \begin{pmatrix}
		\hat{\rho}_A + \hat{\rho}_B \\
		\hat{\rho}_A - \hat{\rho}_B
	\end{pmatrix}.
\end{equation}
Eigenvalues ${\rm Re}(\sigma_{\rho})>0$ indicate density instabilities related to (general) phase separation phenomena. We can distinguish different types of phase separation by monitoring the corresponding eigenvectors $\bm{v}_{\rho} = (\hat{\rho}_A + \hat{\rho}_B, \hat{\rho}_A - \hat{\rho}_B)^{\rm T}$. Of particular importance is the angle $\alpha = \arccos(\bm{v}_{\rho}\cdot (1,0)^{\rm T})$, which specifies the direction of the eigenvector with largest ${\rm Re}(\sigma_{\rho})$. An angle $\alpha=0$ indicates pure total density fluctuations, and, thus, symmetric clustering. This small-wavenumber instability is related to MIPS \cite{speck_loewen_2014_effective_Cahn-Hilliard_ABP,Bialke_2013_microscopic_theory_phase_seperation,speck_2015_dynamical_mean_field_phase_separation}.
In contrast, full symmetric demixing occurs when $\alpha=\pm\pi/2$. Finally asymmetric clustering (i.e., partial demixing) of predominantly species $A$ ($B$) corresponds to $0<\alpha<\pi/2$ ($-\pi/2 < \alpha < 0$).

For reciprocal systems with $g'_{AB}=g'_{BA}=\kappa$, the stability matrix is diagonal,
\begin{equation}
	\label{eq:stability_matrix_coase-grained_density_reciprocal}
	\bm{\mathcal{M}}^{\rm rec}_{\rm ad} = \tfrac{V}{2} \, k^2 \begin{pmatrix}
		\frac{V-2\,z}{g'- D_{\rm r} + \kappa} & 0 \\
		0 & \frac{V}{g'- D_{\rm r} - \kappa}
	\end{pmatrix}
\end{equation}
with $V={\rm Pe}-2\,z$ (and $D_{\rm t}=0$ \footnote{Translational diffusion damps eigenvalues with $\sim -D_{\rm t}\,k^2$ \cite{SM}.}). As expected, we then observe only symmetric density dynamics affecting both species equally. Specifically, outside the flocking regime (with $\vert \kappa \vert $ small), we find MIPS-like symmetric clustering ($\alpha=0$). Demixing ($\alpha=\pm \pi/2$) is predicted only for large, negative $\kappa$ within the antiflocking-regime. However, then, the assumption of small polarization fields does not hold \cite{SM}, and one needs to consider the full six-dimensional problem.

Non-reciprocal orientational couplings destroy symmetric phase separation. To see this, we choose fully antisymmetric couplings, $g'_{AB} = -g'_{BA} = \delta$, yielding
\begin{equation}
	\bm{\mathcal{M}}^{\rm nr}_{\rm ad} = \tfrac{V\, k^2}{2(\delta^2 + (D_{\rm r}-g')^2)}  \begin{pmatrix}
		(g'- D_{\rm r}) (V-2\,z) & V\,\delta \\
		-(V-2\,z)\,\delta & (g'- D_{\rm r})V
	\end{pmatrix}.
\end{equation}
The eigenvectors now predict asymmetric clustering with $\alpha \neq 0, \pm \pi/2$, depending on the sign and magnitude of $\delta$. The clustering angle is shown in Fig.~\ref{fig:stability_diagram_with_snapshots}(f). For negative $\delta$, asymmetric $B$-clustering is predicted. Increasing $\delta$ gradually decreases the degree of $B$-clustering, first, to symmetric clustering at $\delta=0$ and then to asymmetric $A$-clustering for positive $\delta$. Notably, the asymmetric clustering is solely induced by $\delta$; it does not occur for reciprocal polar alignment (nor in the passive case \cite{SM}). 
This is also seen in particle simulations without repulsion, see \cite{kreienkamp_klapp_PRE}.
Corresponding BD snapshots of the asymmetric clustering are shown in Figs.~\ref{fig:stability_diagram_with_snapshots}(d),(e).
Simulation movies \cite{SM} in this regime reveal ``chasing'' behaviors familiar from other non-reciprocal off-lattice systems \cite{You_Baskaran_Marchetti_2020_pnas,saha_scalar_active_mixtures_2020,agudo-canalejo_golestanian_2019_active_phase_separation_chemically_interacting_particles,mandal_sollich_2022_robustness_travelling_states_generic_non-reciprocal_mixture,chiu_Omar_2023_phase_coexistence_implications_violating_Newtons_third_law}.

For very strong non-reciprocity, translational diffusion fully suppresses phase separation, yielding a homogeneous disordered state.

BD simulations of the Langevin equations \eqref{eq:Langevin_eq} are qualitatively consistent
with predictions from the continuum level (see marker points in Fig.~\ref{fig:stability_diagram_with_snapshots}(b)). A full description is given in \cite{kreienkamp_klapp_PRE}.

\begin{figure}
	\includegraphics[width=1\linewidth]{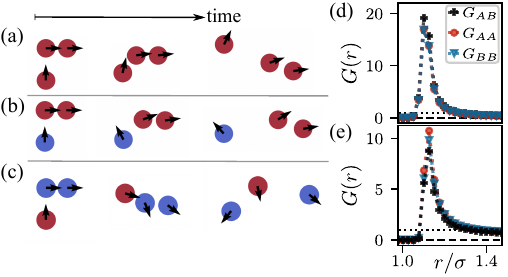}
	\caption{\label{fig:sketch_asymmetric_clustering}
Microscopic origin and resulting asymmetric clustering. (a)-(c) Illustration of particle motion leading to asymmetric clustering of species $A$ ($\delta,g>0$). Particles of species $A$ ($B$) are colored in red (blue). Small two-particle $A$-clusters survive (a,b), while $B$-clusters are destabilized (c) upon approach of a third particle. Pair correlation functions for (d) $g_{AB}=g_{BA}=9$, \mbox{(e) $g_{AB}=-g_{BA}=9$} (and $g=3$). Data represent time averages between $0.5$ and $1\,\tau$ after initialization.}
\end{figure}

\textit{Microscopic origin of asymmetric clustering.}---The origin of asymmetric clustering can be understood on a microscopic level. We consider the case $\delta,g>0$, such that
$A$ tends to orient along $B$ (and $A$) while $B$ wants to orient opposite to $A$ (and along $B$).
In Fig.~\ref{fig:sketch_asymmetric_clustering} we illustrate the evolution of a small ``cluster'' involving two coherently moving $A$- or $B$-particles upon approach of a third particle. 
If the approaching particle is from the same species, it either joins the cluster; or at least, does not significantly disturb its motion [case (a)]. If a $B$-particle approaches an $A$-cluster, it quickly reorients into the opposite direction (since $g_{BA}<0$) and thereby tends to move away [case (b)]. Thus, the $B$-particle does not disturb the $A$-cluster. In contrast, if an $A$-particle approaches a $B$-cluster, it tends to orient along the cluster's direction ($g_{AB}>0$). This disturbs the coherent motion of the $B$-particles and the $B$-cluster is destabilized [case (c)].

\textit{Microscopic fluctuation analysis.}---We now compare the predicted hydrodynamic phase behavior to the microscopic one on a quantitative level.

The pair correlation function $G_{ab}(r)$ measures the distribution of $a$-particles around a particle of species $b$ at distance $r$ \cite{SM}. Hence, it contains information regarding the (a)symmetry of clustering in particle simulations. In Fig.~\ref{fig:sketch_asymmetric_clustering}(d,e), we plot $G_{ab}(r)$ shortly after initialization from a disordered configuration.
In the reciprocal case, we always find $G_{AA}=G_{BB}$, while $G_{AB}$ may be smaller or larger depending on the ratio $g/\kappa$. In contrast, asymmetric clustering is characterized by $G_{AA}\neq G_{BB}$. In particular, for $\delta>0$, $G_{AA}>G_{BB}$ indicates the preference of $A$-clustering.

We now use the short-time correlations for a systematic analysis of density fluctuations \cite{chen_forstmann_1992_demixing_binary_Yukawa_fluid, chen_forstmann_1992_phase_instability_fluid_mixtures,range_klapp_2006_pair_formation_ferrocolloid_mixtures}.
We consider long-wavelength fluctuations of the total density, $\delta\hat{\rho}(k) = \delta \hat{\rho}_A(k) + \delta \hat{\rho}_B(k)$, 
the composition $\delta\hat{c}(k) = \delta \hat{\rho}_A(k) - \delta \hat{\rho}_B(k)$, and mixtures of these. Their magnitude is given by the structure factors $S_{ij}(k)=\langle \delta\hat{i}(k) \, \delta\hat{j}(k) \rangle$ ($i,j=\rho,c$) that can be computed as Fourier transforms of $G_{ab}(r)$ \cite{SM,chen_forstmann_1992_demixing_binary_Yukawa_fluid, chen_forstmann_1992_phase_instability_fluid_mixtures}. The structure factors $S_{ij}(k)$ form the symmetric matrix $\bm{\mathcal{S}}$. In the following, we focus on the limit $k\to 0$, which turns out to be most relevant. 
If the homogeneous system becomes unstable, correlations of density fluctuations are expected to diverge. Thus, an instability is signaled by the divergence of one eigenvalue $\lambda_{1/2}$ of $\bm{\mathcal{S}}$, or equivalently, a vanishing of its inverse, $\lambda_{1/2}^{-1}$. The dominant character of the instability, i.e., the type of phase separation, is indicated by the corresponding eigenvector. As in the continuum analysis, the direction of the eigenvector in the $\delta\hat{\rho}$\,--\,$\delta\hat{c}$-plane is quantified by the angle $\alpha$ and indicates symmetric clustering, symmetric demixing or asymmetric clustering. 

\begin{figure}
	\includegraphics[width=1\linewidth]{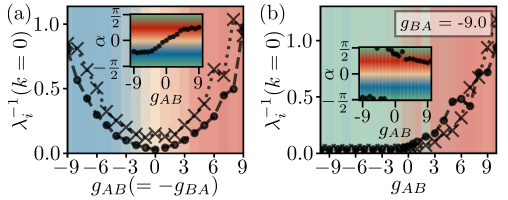}
	\caption{\label{fig:eigenvalues_structure_factor} (Inverse) eigenvalues $\lambda_{1/2}^{-1}$ of $\bm{\mathcal{S}}(k\rightarrow 0)$ and predicted angle $\alpha$ (inset) from BD simulations. 
	\mbox{(a) $g_{AB}=-g_{BA}$.} (b) $g_{BA}=-9$ (and $g=3$). Data represent time averages between $4.5$ and $5\,\tau$ after initialization.} %
\end{figure}

Eigenvalues and corresponding $\alpha$ from BD simulations are shown in Fig.~\ref{fig:eigenvalues_structure_factor}. In the reciprocal case, $\delta=0$ in Fig.~\ref{fig:eigenvalues_structure_factor}(a), the fluctuation analysis indicates a symmetric clustering instability with $\lambda_1^{-1}\approx 0$ and $\alpha\approx 0$. Moving into the non-reciprocal regime by increasing $|\delta |$, 
$\lambda_{1,2}^{-1}$ become non-zero. Thus, density fluctuations are strong, yet not divergent anymore. At the same time, $\alpha$ continuously changes and now indicates asymmetric clustering.
In Fig.~\ref{fig:stability_diagram_with_snapshots}(b), we move along a horizontal path from reciprocal antiflocking towards the non-reciprocal $A$-clustering regime. At the beginning, $\lambda_{1/2}^{-1}$ are close to zero.
Together with $\alpha\approx \pm\pi/2$ this means that the reciprocal antiflocking state is associated to a demixing instability. With increasing non-reciprocity, asymmetric $A$-clustering becomes more and more dominant.

These observations conform with visual inspection of snapshots [Fig.~\ref{fig:stability_diagram_with_snapshots}(c)-(e)] and predictions on the continuum level. Indeed, the agreement between the different levels of description (BD simulations versus continuum) holds also quantitatively. This is seen in Fig.~\ref{fig:stability_diagram_with_snapshots}(f), where we directly compare the angles $\alpha$ from the two types of calculations.

\textit{Towards larger coupling strengths.}---All of the above results have been obtained in the weak-intraspecies-coupling regime, where the dominant eigenvalues are real-valued essentially everywhere.
This changes when we increase the intraspecies couplings. Beyond the critical flocking line $g' = D_{\rm r}$, the largest $k=0$-eigenvalue is always positive. Its imaginary part becomes nonzero as soon as species have opposite goals ($g_{AB}\,g_{BA}<0$). Within this regime, we find exceptional points (eigenvalue coalescence with parallel eigenvectors, see \cite{SM}) that have been related to parity-time symmetry breaking of the dynamics in simpler non-reciprocal systems \cite{You_Baskaran_Marchetti_2020_pnas,fruchart_2021_non-reciprocal_phase_transitions,saha_scalar_active_mixtures_2020}. A more detailed analysis of this phenomenon and its interplay with density dynamics will be given in a future paper.

\textit{Conclusion.}---We demonstrate how non-reciprocal orientational couplings affect the density dynamics in active mixtures. In particular, our results reveal asymmetric clustering and, thus, partial demixing in otherwise fully symmetric systems. This behavior occurs on the hydrodynamic level of description as well as on the microscopic scale (discussed in more detail in \cite{kreienkamp_klapp_PRE}). The remarkable consistency is not clear {\it ad hoc} \cite{dinelli_tailleur_2023_non-reciprocity_across_scales}. 

The behavior found here is in stark contrast to equilibrium mixtures where demixing rather results from large interspecies attraction \cite{wilding_nielaba_1998_liquid-vapor_phase_behavior_symmetrical_binary_fluid_mixture} or different particle shapes \cite{adams_fraden_1998_entropically_driven_phase_transitions_mixtures_rods_spheres}. In active systems, demixing has already been shown to result from conservative interactions \cite{agudo-canalejo_golestanian_2019_active_phase_separation_chemically_interacting_particles,duan_mahault_2023_dynamical_pattern_formation_without_self-attraction} or differences in translational parameters such as diffusion constants \cite{weber_frey_2016_binary_mixture_different_diffusivities_demix} and active speeds 
\cite{Adhikary_Santra_2022_pattern_formation_phase_transition_binary_mixture,kolb_2020_active_binary_mixture_slow_fast,ilker_joanny_2020_phase_separation_mixtures_different_temperatures}. In contrast, here, the only difference between particles lies in their non-reciprocal interspecies torques.

Our results could, in principle, be tested in mixtures of Quincke rollers \cite{bricard_bartolo_2013_emergence_macroscopic_directed_motion,maity_morin_2023_spontaneous_demixing_binary_colloidal_flocks} or engineered in robotic experiments \cite{chen_zhang_2024_emergent_chirality_hyperuniformity_active_mixture_NR}. 
Future work should focus on the dynamics in the presence of exceptional transitions as well as on thermodynamic implications \cite{loos_klapp_2020_irreversibility_heat_non-reciprocal_interactions,suchanek_loos_2023_irreversible_fluctuations_emergence_dynamical_phases,suchanek_loos_2023_time-reversal_PT_symmetry_breaking_non-Hermitian_field_theories}.

\begin{acknowledgments}
This work was funded by the Deutsche Forschungsgemeinschaft (DFG, German Research Foundation) -- Projektnummer 163436311 (SFB 910) and Projektnummer 517665044. We thank Jan Meibohm for critical reading.
\end{acknowledgments}

\input{main.bbl}

\end{document}


\title{Supplemental Material to ``Non-reciprocal alignment induces asymmetric clustering in active
	mixtures''}

\author{Kim L. Kreienkamp}
\email{k.kreienkamp@tu-berlin.de}
\author{Sabine H. L. Klapp}
\email{sabine.klapp@tu-berlin.de}
\affiliation{%
 Institut f\"ur Theoretische Physik, Technische Universit\"at Berlin
}%



\maketitle

In this supplemental, we provide additional background on our models and methods of analysis. In particular, we introduce in detail the microscopic Langevin equations as well as the associated continuum model that is used for the linear stability analysis and characterization of instabilities on the mean-field level. Further, we define the pair correlation functions based on particle trajectories. Finally, we present the connection between the pair correlations and the structure factor matrix, which is used to quantify the degree of (a)symmetrical clustering. 

\tableofcontents

\section{\label{app:microscopic_model}Microscopic model}
We consider a two-dimensional system of active particles comprising two species $a=A,B$. The binary mixture contains $N = N_A + N_B$ particles that are located at positions $\bm{r}_{\alpha}$ (with $\alpha=i_a = 1,...,N_a$) and move like active Brownian particles (ABP). They are subject to an additional torque due to orientational couplings. They self-propel with velocity $v_0$ in the direction $\bm{p}_{\alpha}(t)=(\cos\,\theta_{\alpha}, \sin\,\theta_{\alpha})^{\rm{T}}$, where $\theta_{\alpha}$ is the polar angle. The overdamped Langevin equations (LE), governing the dynamics, are given by
\begin{subequations} \label{eq:Langevin_eq}
	\begin{align}
		\dot{\bm{r}}_{\alpha}(t) &= v_0\,\bm{p}_{\alpha}(t) + \mu_{r} \sum_{\beta\neq\alpha} \bm{F}_{\rm{rep}}^{\alpha}(\bm{r}_{\alpha},\bm{r}_{\beta}) + \sqrt{2\,D'_{\rm t}}\,\bm{\xi}_{i}^a(t) \label{eq:Langevin_r}\\
		\dot{\theta}_{\alpha}(t) &= \mu_{\theta} \sum_{\beta\neq\alpha} \mathcal{T}_{\rm al}^{\alpha}(\bm{r}_{\alpha},\bm{r}_{\beta},\theta_{\alpha},\theta_{\beta}) + \sqrt{2\,D'_{\rm r}}\,\eta_{i}^a(t) \label{eq:Langevin_theta},
	\end{align}
\end{subequations}
where the sums over particles $\beta=j_b=1,...,N_b$ couple the dynamics of particle $\alpha$ to the position and orientation of all other particles of both species $b=A,B$. 

The translational LE \eqref{eq:Langevin_r} contains the repulsive force $\bm{F}_{\rm{rep}}^{\alpha} = - \sum_{\beta\neq\alpha} \nabla_{\alpha} U(r_{\alpha\beta})$ between hard disks, derived from the Weeks-Chandler-Andersen (WCA) potential \cite{weeks_1971_Weeks-Chandler-Andersen_potential} 
{\small
	\begin{equation}
		\label{eq:WCA_potential}
		U(r_{\alpha\beta}) = \begin{cases}
			4\epsilon \left[\left( \frac{\sigma}{r_{\alpha\beta}}\right)^{12} - \left(\frac{\sigma}{r_{\alpha\beta}}\right)^6 + \frac{1}{4} \right], \ {\rm if} \ r_{\alpha\beta}<r_{\rm c}\\
			0, \ {\rm else}
		\end{cases} ,
	\end{equation}
}%
where $r_{\alpha\beta} = \vert \bm{r}_{\alpha\beta}\vert = \vert \bm{r}_{\alpha}-\bm{r}_{\beta} \vert$. The characteristic energy scale of the potential is given by $\epsilon$. The cut-off distance is $r_{\rm c }=2^{1/6}\,\sigma$. The particle diameter $\sigma$ is taken as characteristic length scale, $\ell = \sigma$.

The rotational LE \eqref{eq:Langevin_theta} involves the torque given by
\begin{equation}
	\label{eq:torque}
	\mathcal{T}_{\rm al}^{\alpha}(\bm{r}_{\alpha}, \bm{r}_{\beta}, \theta_{\alpha}, \theta_{\beta}) = k_{ab}\, \sin(\theta_{\beta}-\theta_{\alpha}) \, \Theta(R_{\theta}-r_{\alpha\beta}).
\end{equation}
Here, $k_{ab}$ denotes its strength and can be positive or negative. The step function with $\Theta(R_{\theta}-r_{\alpha\beta})=1$ if $r_{\alpha\beta}<R_{\theta}$ and zero otherwise, ensures that only particles within radius $R_{\theta}$ interact via the torque. Particles of species $a$ tend to orient parallel (align) or antiparallel (antialign) with neighboring particles of species $b$ when $k_{ab}>0$ or $k_{ab}<0$, respectively. \textit{Reciprocal} couplings are defined by the choice $k_{AB} = k_{BA}$. Then, particles of species $A$ align (or antialign) with particles of species $B$ in the same way as particles of species $B$ with particles of species $A$. \textit{Non-reciprocal} orientational couplings have $k_{AB}\neq k_{BA}$.

Both the position and orientation of the particles are subject to thermal noise, modeled as Gaussian white noise processes $\bm{\xi}_{\alpha}$(t) and $\eta_{\alpha}(t)$ of zero mean and variances $\langle \xi_{\alpha,k}(t) \xi_{\beta,l}(t') \rangle = \delta_{\alpha\beta}\,\delta_{kl} \,\delta(t-t')$ and $\langle \eta_{\alpha}(t) \eta_{\beta}(t') \rangle = \delta_{\alpha\beta}\,\delta(t-t')$, respectively. The (Brownian) time a (passive) particle needs to travel over its own distance is $\tau=\sigma^2/D'_{\rm t}$, which we take as characteristic time scale. The mobilities fulfill the Einstein relation and are connected to thermal noise via $\mu_{r} = \beta\,D'_{\rm t}$ and $\mu_{\theta} = \beta \, D'_{\rm r}$, where $\beta^{-1}=k_{\rm B}\,T$ is the thermal energy with Boltzmann's constant $k_{\rm B}$ and temperature $T$.

We introduce the following dimensionless parameters: the average area fractions $\Phi_a = \rho_0^a\,\pi\,\ell^2/4$ of species $a$ with (number) density $\rho_0^a = N_a/L^2$, the reduced orientational coupling parameter $g_{ab} = k_{ab}\,\mu_{\theta}\,\tau,$ and the P\'eclet number ${\rm Pe} = v_0\,\tau/\ell$, which quantifies the persistence of the motion of particles.

We perform numerical Brownian Dynamics (BD) simulations of the LE \eqref{eq:Langevin_eq} to study the emerging dynamical structures in our system. In our simulations, we use a fixed combined average area fraction $\Phi = 0.4$, where $\Phi_A=\Phi_B=0.2$, and P\'eclet number ${\rm Pe}=40$, while varying the orientational couplings strengths $g_{ab}$. We simulate $N=5000$ particles, with equal particle numbers $N_A = N_B = 2500$ of both species, in a $L\times L$ box subjected to periodic boundary conditions. We use the particle diameter $\sigma$ as characteristic length scale, $\ell = \sigma = 1$, and the time unit as $\tau = \sigma^2/D'_{\rm t} = 1$. The repulsive strength is chosen to be $\epsilon^{*} = \epsilon / (k_{\rm B}\,T) = 100$, where the thermal energy is set to be the energy unit ($k_{\rm B}\,T=1$). The diffusion constants are then given by $D'_{\rm t} = 1 \, \ell^2/\tau$ and $D'_{\rm r} = 3 \cdot 2^{-1/3}/\tau$. The cut-off distance for the torque is chosen to be $R_{\theta}=2\,\ell$. The simulations are performed by initializing the system in a random configuration, integrating the equations of motions using an Euler-Mayurama algorithm, and letting the system evolve into its steady state before measuring quantities for phase characterization. To this end, we employ a timestep of $\Delta t = 10^{-5}\,\tau$ until the simulations have lasted for $120\,\tau$.

\begin{figure*}
	\includegraphics[width=\linewidth]{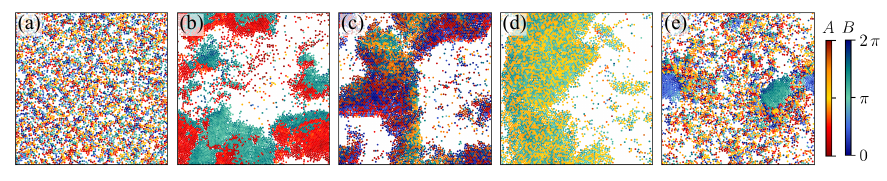}
	\caption{\label{fig:snapshots_BD_simulations}Snapshots of Brownian dynamics simulations for (a) $g_{AB}=-g_{BA}=20$ (disorder). (b) $g_{AB}=g_{BA}=-9$ (antiflocking). (c) $g_{AB}=g_{BA}=1$ (MIPS). (d) $g_{AB}=g_{BA}=9$ (flocking and phase separation). (e) $g_{AB}=-g_{BA}=-9$ (asymmetric clustering of species $B$). Other parameters are $g_{AA}=g_{BB}=3$, ${\rm Pe}=40$, $\Phi=0.4$, and $D_{\rm r}=2 \cdot 3^{-1/3}$.}
\end{figure*}

Snapshots of the Brownian dynamics simulations for different parameter combinations are shown in Fig.~\ref{fig:snapshots_BD_simulations}. 

\section{\label{app:continuum_model}Continuum model}
For the derivation of the continuum model associated to LE \eqref{eq:Langevin_eq}, we refer to \cite{kreienkamp_klapp_2022_clustering_flocking_chiral_active_particles_non-reciprocal_couplings,kreienkamp_klapp_PRE}. The final equations comprise the continuity equation for the densities $\rho^a(\bm{r},t)$,
\begin{equation}
	\label{eq:continuum_eq_density}
	\partial_t \rho^a +  \nabla \cdot \bm{j}_{a} = 0
\end{equation}
with flux
\begin{equation}
	\label{eq:continuity_flux}
	\bm{j}_a = v^{\rm eff}(\rho)\, \bm{w}^a -  D_{\rm t}\,\nabla\,\rho^a .
\end{equation} 
The flux involves the polarization densities $\bm{w}^a(\bm{r},t) = \rho^a\,\bm{P}^a$ with $\bm{P}^a(\bm{x},t)$ being the polarization field, measuring the overall orientation of particles at a certain position. The polarization density $\bm{w}^a$ evolves according to
{\small 
	\begin{equation}
		\label{eq:continuum_eq_polarization}
		\begin{split}
			\partial_t & \bm{w}^a  \\
			=& - \frac{1}{2} \, \nabla \,\big(v^{\rm eff}(\rho)\, \rho^a\big)  - D_{\rm r }\, \bm{w}^{a} + \sum_b g'_{ab} \, \rho^a\, \bm{w}^b +  D_{\rm t}\,\nabla^2\,\bm{w}^a + \frac{v^{\rm eff}(\rho)}{16\,D_{\rm r}} \, \nabla^2\,\Big(v^{\rm eff}(\rho)\,\bm{w}^a \Big) - \sum_{b,c} \frac{ g'_{ab}\,g'_{ac}}{2\,D_{\rm r}} \, \bm{w}^a \, (\bm{w}^b \cdot \bm{w}^c) \\
			&- \frac{z}{16\,D_{\rm r}} \, \nabla\rho \cdot \big[\nabla \big(v^{\rm eff}(\rho)\,\bm{w}^a\big) - \nabla^* \big(v^{\rm eff}(\rho)\,\bm{w}^{a*} \big) \big] + \sum_b \frac{g'_{ab}}{8 \,D_{\rm r}} \Big[ \bm{w}^b \cdot \nabla \big(v^{\rm eff}(\rho)\, \bm{w}^a\big) + \bm{w}^{b*} \cdot \nabla \big(v^{\rm eff}(\rho)\, \bm{w}^{a*}\big) \\
			&\qquad \qquad \qquad   -2 \, \Big\{ v^{\rm eff}(\rho)\,\bm{w}^a \cdot \nabla \bm{w}^b  +\bm{w}^b \, \nabla \cdot \big(v^{\rm eff}(\rho)\,\bm{w}^a\big) - v^{\rm eff}(\rho)\, \bm{w}^{a*} \cdot \nabla \bm{w}^{b*} - \bm{w}^{b*}\,  \nabla \cdot \big( v^{\rm eff}(\rho)\, \bm{w}^{a*}\big) \Big \}  \Big] .
		\end{split}
\end{equation} }%
The density flux $\bm{j}_a$ given in Eq.~\eqref{eq:continuity_flux} comprises that particles of species $a$ move in space due to their self-propulsion in the direction $\bm{w}^a$. Importantly, the self-propulsion velocity is not constant but particles slow down in high-density regions. This is reflected in the density-dependent velocity 
\begin{equation}
	v^{\rm eff}(\rho)={\rm Pe}- z\,\rho
\end{equation}
with $\rho=\sum_b\rho^b$. The flux further comprises translational diffusion. 
The evolution of the polarization density $\bm{w}^a$, given by Eq.~\eqref{eq:continuum_eq_polarization}, has various contributions: Particles tend to swim (with increasing speed) towards low-density regions (first term on right-hand side), the polarization decays due to rotational diffusion (second term), and orientations of all particles are coupled (third term). The remaining terms are diffusional and non-linear contributions, which smooth out low- and high-polarization regions.

In Eq.~\eqref{eq:continuum_eq_polarization}, we have introduced $\bm{w}^*=(w_y, -w_x)^{\rm T}$ and $\nabla^*=(\partial_y, -\partial_x)^{\rm T}$. We non-dimensionalized the equations with a characteristic time scale $\tau$ and a characteristic length scale $\ell$. Further, particle and polarization densities of species $a$ are scaled with the average particle density $\rho_0^a$. There are five remaining dimensionless control parameters. These are the P\'eclet number ${\rm Pe}=v_0\,\tau/\ell$, the velocity-reduction parameter $z = \zeta\,\rho_0^a\,\tau/\ell$ due to the environment, the translational diffusion coefficient $D_{\rm t}=D'_{\rm t}\,\tau/\ell^2$, the rotational diffusion coefficient $D_{\rm r}=D'_{\rm r}\,\tau$, and the relative orientational coupling parameter $g'_{ab} = k_{ab}\,\mu_{\theta}\,R_{\theta}^2\,\rho_0^b\,\tau/2$. The five control parameters are summarized in Table \ref{tab:control_parameters}.

\begin{table}
	\begin{tabular}{lll}
		\hline
		parameter & definition & description\\
		\hline
		${\rm Pe}$ & $v_0\,\tau/\ell$ & P\'eclet number\\
		$z$ & $\zeta\,\rho_0^a\,\tau/\ell$ & particle velocity-reduction\\
		$D_{\rm t}$ & $D'_{\rm t}\,\tau/\ell^2$ & translational diffusion\\
		$D_{\rm r}$ & $D'_{\rm r}\,\tau$ & rotational diffusion\\
		$g'_{ab}$ & $k_{ab}\,\mu_{\theta}\,R_{\theta}^2\,\rho_0^b\,\tau/2$ & orient. coupling strength\\
		\hline
	\end{tabular}
	\caption{\label{tab:control_parameters}The five control parameters in the non-dimensionalized continuum Eqs.~\eqref{eq:continuum_eq_density} -- \eqref{eq:continuum_eq_polarization} with characteristic time and length scales, $\tau$ and $\ell$. The average density is $\rho_0 = \sum_a \rho_0^a$ with single-species density $\rho_0^a$.}
\end{table}

\subsection{\label{sapp:parameter_choice_continuum}Parameter choice with respect to particle-based model}
In our continuum model, most parameters can be directly adopted from the considered particle simulation parameters. These include the P\'eclet number, ${\rm Pe}=40$, and the rotational diffusion constant, $D_{\rm r} = D'_{\rm r}\,\tau = 3 \cdot 2^{-1/3}$. The area fraction in particle simulations, $\Phi = 0.4$, corresponds to the number density $\rho_0 = 2\, \rho_0^a = 4/\pi\,\Phi$, where $\rho^a_0 = 2/\pi\,\Phi$.  The orientational couplings in continuum simulations ($g'_{ab}$) are related to those in the particle simulations ($g_{ab}$) via $g_{ab}'=0.51\,g_{ab}$, given $R_{\theta}=2\,\ell$. We consider a case with fixed weak intraspecies coupling strengths, $g_{AA}=g_{BB}=3$, while the interspecies coupling strengths $g_{AB}$ and $g_{BA}$ are chosen independently.  However, there are two parameters that require special attention: the velocity reduction parameter, $\zeta$, and the translational diffusion constant, $D_{\rm t}$. For details regarding the parameter choice, see \cite{kreienkamp_klapp_PRE}. We choose $D_{\rm t}=9$.  Further, we can directly obtain the non-dimensionalized velocity reduction parameter $z = 57.63 \, \rho^a_0 \, \tau/\ell = 0.37 \, {\rm Pe}/\rho_0^{\rm con}$ with $\rho_0^{\rm con}=1$ from particle simulations. This velocity reduction parameter places the system well within the MIPS instability region for a wider range of alignment strengths \cite{kreienkamp_klapp_2022_clustering_flocking_chiral_active_particles_non-reciprocal_couplings}. 

\begin{figure*}
	\includegraphics[width=\linewidth]{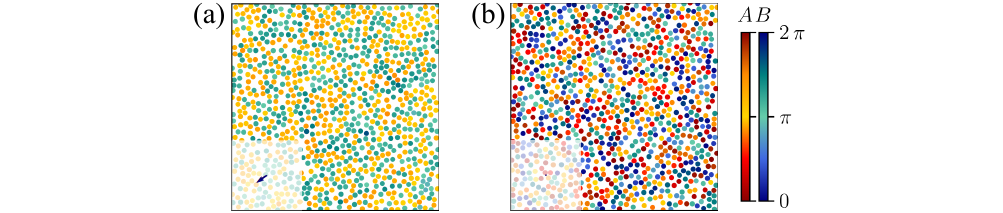}
	\caption{\label{fig:snapshots_BD_simulations_passive}Snapshots of Brownian Dynamics simulations of passive particles [$v_0=0$ in Eqs.~(1a),(1b)]. Interactions between the particles include steric repulsion and (non-)reciprocal alignment couplings. (a) $g_{AB}=g_{BA}=9$ (reciprocal). (b) $g_{AB}=-g_{BA}=9$ (non-reciprocal). The arrows in the lower left corner indicate the polarization (vector) of individual species. They overlap in case of reciprocal alignment [(a)]. Other parameters are $g_{AA}=g_{BB}=3$, ${\rm Pe}=40$, $\Phi=0.4$, $N=1000$, and $D_{\rm r}=2 \cdot 3^{-1/3}$.}
\end{figure*}

\subsection{\label{sapp:passive_limit_continuum}The effect of steric repulsion in the passive limit}
The passive limit is defined by vanishing velocity of particles, i.e., $v_0=0$ on the microscopic level [Eqs.~(1a),(1b)] and $v^{\rm eff} = 0$ on the continuum level [Eqs.~(4)-(6)]. In the following, we relate the microscopic and continuum descriptions in the passive limit with a special focus on steric repulsion between particles.

On the microscopic level [Eqs.~(1a),(1b)], passive particles with $v_0 = 0$ interact via steric repulsion and \mbox{(non-)}reciprocal alignment couplings. In case of a single aligning species, the so-defined system corresponds to a well-studied model, namely, the classical Heisenberg fluid \cite{hoye_stell_1976_configurationally_disordered_spin_systems,hemmer_imbro_1977_ferromagnetic_fluids,oukouiss_baus_phase_diagrams_classical_heisenberg_fluid}. 
The phase behavior of the classical Heisenberg fluid depends on the density and strength of ferromagnetic couplings as compared to temperature. Roughly speaking, these phases can be summarized as following. For small densities and temperatures, the Heisenberg fluid phase separates into a (paramagnetic) gas phase without order and a liquid phase (either with or without ferromagnetic order). For larger temperatures, the system does not phase separate and is either in a paramagnetic or ferromagnetic fluid phase. Large densities lead to solid phases \cite{oukouiss_baus_phase_diagrams_classical_heisenberg_fluid}.

Qualitatively, we can compare the phase separation in our binary mixture to the one in a classical Heisenberg fluid. The overall area fraction in our system is $\Phi=0.4$. Inspecting snapshots of \textit{passive} systems (with $v_0=0$) at this area fraction (Fig.~\ref{fig:snapshots_BD_simulations_passive}), we clearly see that particles do not form any significant clusters regardless of the considered alignment couplings. In case of reciprocal alignment [Fig.~\ref{fig:snapshots_BD_simulations_passive}(a)], orientations of particles are correlated over long distances. Therefore, the phase of the reciprocally aligning system is reminiscent of the ferromagnetic fluid phase of the classical Heisenberg fluid. In the non-reciprocal case [Fig.~\ref{fig:snapshots_BD_simulations_passive}(b)], particle orientations are correlated only over short distances and the polarization in the system is zero. However, most importantly for modeling the effect of steric repulsion, Brownian Dynamics simulations reveal that particle positions are only weakly correlated (i.e., spatial correlations are, at most, short-ranged). The average particle density field would therefore correspond to a homogeneous density field on the continuum level.

On the continuum level [Eqs.~(4)-(6)], the density dynamics in the passive limit reduces to two uncoupled, diffusing density fields. The polarization fields, on the other hand, still comprise couplings between the own and other species' polarization fields. However, as a consequence of the mean-field assumption, the polarization fields do not affect the density dynamics. Thus, the resulting hydrodynamic density fields are homogeneous. This corresponds sufficiently well to the observations on the particle level for the passive case at the volume fraction of interest.

In principle, one could describe short-range steric repulsion also for the passive case on the continuum level. This could be done, e.g., in the framework of classical dynamical density functional theory, where correlation effects are handled via a free-energy functional \cite{marconi_tarazona_1999_dynamic_density_functional_theory_fluids,archer_2004_dynamical_density_function_theory}, or, even simpler, by density-dependent diffusion coefficients \cite{felderhof_1978_diffusion_interaction_Brownian_particles}. Nevertheless, the resulting density dynamics would not show any clustering effects, consistent with the associated particle system. Therefore, our (strongly simplified) density dynamics in the passive limit does not affect our predictions for the active mixture.

\section{\label{app:linear_stability_analysis}Linear stability analysis}
\subsection{Stability matrix of full dynamics}
We analytically determine the linear stability of the homogeneous disordered state characterized by a uniform density and zero polarization for both species $a=A,B$, i.e., $(\rho^a, \bm{w}^a)=(1,\bm{0})$. Remember that the continuum Eqs.~\eqref{eq:continuum_eq_density} - \eqref{eq:continuum_eq_polarization} are already scaled with the mean density $\rho_0^a$. We consider perturbations of the disordered state involving all wave vectors $\bm{k}$, expressed as plane waves with a (complex) growth rate $\sigma(k)$ and amplitudes $\hat{\rho}^a(k)$ and $\bm{\hat{w}}^a(k)$ \cite{kreienkamp_klapp_PRE}. Since we investigate the stability of the isotropic disordered base state, the growth rate $\sigma$ depends only on the wave number $k=\vert \bm{k} \vert$.

We are interested in perturbations in the combined field quantities $\rho^A+\rho^B$, $\rho^A-\rho^B$, $\bm{w}^A+\bm{w}^B$, and $\bm{w}^A-\bm{w}^B$. Starting from the continuum Eqs.~\eqref{eq:continuum_eq_density} - \eqref{eq:continuum_eq_polarization}, linearization with respect to the perturbation leads to an eigenvalue equation
\begin{equation}
	\label{eq:eigenvalue_equation}
	\sigma(k) \, \bm{v}(k) = \bm{\mathcal{M}}(k) \cdot \bm{v}(k)
\end{equation}
for each $k$.
The eigenvector $\bm{v}(k) = (\hat{\rho}^A+\hat{\rho}^B, \, \hat{\rho}^A-\hat{\rho}^B,\, \hat{w}_x^A + \hat{w}_x^B,\, \hat{w}_y^A + \hat{w}_y^B,\,  \hat{w}_x^A - \hat{w}_x^B,\, \hat{w}_y^A - \hat{w}_y^B)^{\rm T}$ contains the possible perturbations of the particle densities and the two components of the polarization densities. The $6 \times 6$ matrix $\bm{\mathcal{M}}(k)$ is given by
\begin{widetext}
	\begin{equation}
		\label{eq:stability_matrix}
		\bm{\mathcal{M}}(k) = \begin{pmatrix}
			-D_{\rm t}\,k^2 & 0 & -i\,k_x\,V & -i\,k_y\,V & 0 & 0\\
			0 & -D_{\rm t}\,k^2 & 0 & 0 & -i\,k_x\,V & -i\,k_y\,V\\
			-\tfrac{i}{2}(V-2\,z)\,k_x & 0 & C_{++}-D_{w} & 0 & C_{+-} & 0 \\
			-\tfrac{i}{2}(V-2\,z)\,k_y & 0 & 0& C_{++}-D_{w} & 0 & C_{+-}  \\
			0 & -\tfrac{i}{2}V\,k_x & C_{-+} & 0 & C_{--} -D_{w} & 0 \\
			0 & -\tfrac{i}{2}V\,k_y & 0 & C_{-+} & 0 & C_{--} -D_{w} \\
		\end{pmatrix},
	\end{equation}
\end{widetext}
where $V = {\rm Pe} - 2\,z$, $D_w =  \mathcal{D}_a \, k^2 + D_r$, and $\mathcal{D}_a = V^2/(16\,D_r) + \mathcal{D}_{\rm t}$. 
The orientation couplings are given by
\begin{equation}
	C_{++} = \tfrac{1}{2}(g'_{AA}+g'_{AB}+g'_{BA}+g'_{BB}),
\end{equation}
\begin{equation}
	C_{+-} = \tfrac{1}{2}(g'_{AA}-g'_{AB}+g'_{BA}-g'_{BB}),
\end{equation}
\begin{equation}
	C_{-+} = \tfrac{1}{2}(g'_{AA}+g'_{AB}-g'_{BA}-g'_{BB}),
\end{equation}
\begin{equation}
	C_{--} = \tfrac{1}{2}(g'_{AA}-g'_{AB}-g'_{BA}+g'_{BB}).
\end{equation}
Solving the eigenvalue Eq.~\eqref{eq:eigenvalue_equation}, we can analytically determine the (complex) growth rates $\sigma(k)$. The real part of the eigenvalues, ${\rm Re}(\sigma)$, determines the actual growth or decay of the perturbations. Non-zero imaginary parts indicate oscillatory behavior. The disordered state is linearly stable when ${\rm Re}(\sigma(k)) < 0$ for all $k$. On the other hand, the disordered state is linearly unstable if ${\rm Re}(\sigma(k)) > 0$ for any $k$. We use the largest value of the six ${\rm Re}(\sigma)$ and corresponding eigenvector to determine the type of emerging dynamics at short times \cite{kreienkamp_klapp_2022_clustering_flocking_chiral_active_particles_non-reciprocal_couplings}.

\subsection{Full non-equilibrium phase characterization}
\label{ssec:linear_stability_analysis_non-equilibirum_phase_characterization}
We use the eigenvalues and the eigenvector corresponding to the largest eigenvalue to characterize the non-equilibrium phases emerging in our system. 

The stability of the disordered (base) state is determined by the real parts of the six eigenvalues, ${\rm Re}(\sigma_i)$. The disordered state is unstable as soon as any eigenvalue has a positive real part at any wave number $k$. To determine the type of emerging dynamics at short times, we follow our earlier work \cite{kreienkamp_klapp_2022_clustering_flocking_chiral_active_particles_non-reciprocal_couplings} and analyze the largest real eigenvalue and corresponding eigenvector. 

In case real parts of the eigenvalues are positive at zero wave number ($k=0$), the instabilities concern the polarization dynamics, that is, \textit{\mbox{(anti-)}flocking}. The reason is that the growth rate at $k=0$ determines the growth or decay of spatially integrated fields. While the polarization field is not a conserved quantity, the particle density is. Hence, at $k=0$ the density-associated growth rates must vanish and all instabilities must be related to polarization dynamics. The type of flocking (parallel or anti-parallel) is indicated by the eigenvector $\bm{v}(k=0)$ corresponding to ${\rm Re}(\sigma(k=0)) > 0$. The largest entry of the eigenvector in $\hat{\bm{w}}^A + \hat{\bm{w}}^B$ predicts (parallel) flocking, while the largest entry in $\hat{\bm{w}}^A - \hat{\bm{w}}^B$ predicts (anti-parallel) \mbox{anti-}flocking.

The density dynamics corresponds to instabilities at finite wave numbers ($k>0$). Here, we consider only the two density-related entries of the (normalized) eigenvector, $\bm{v}_{\rho}=(\hat{\rho}^A+\hat{\rho}^B, \, \hat{\rho}^A-\hat{\rho}^B)^{\rm T}$, at small $k>0$. \textit{Symmetric clustering} corresponds to $\bm{v}_{\rho}=\bm{x}_{\rm con}=(1,0)^{\rm T}$. The angle $\alpha=\arccos(\bm{v}_{\rho} \cdot \bm{x}_{\rm con})$ between $\bm{v}_{\rho}$ and $\bm{x}_{\rm con}$ is approximately $0$. 
\textit{Demixing} corresponds to $\bm{v}_{\rho}$ close to $(0,1)^{\rm T}$ with $\alpha\approx \pm \pi/2$. \textit{Asymmetrical clustering} is quantified by intermediate angles $\alpha$: For asymmetrical clusters of species $A\ (B)$, the angle is $0<\alpha < \pi/2\ (-\pi/2<\alpha < 0)$. The degree of clustering depends on the considered wave number $k$. We are interested in large-wavelength perturbations and therefore consider the limit of small (but finite) wave numbers when characterizing the clustering instabilities.

\begin{figure*}
	\includegraphics[width=\linewidth]{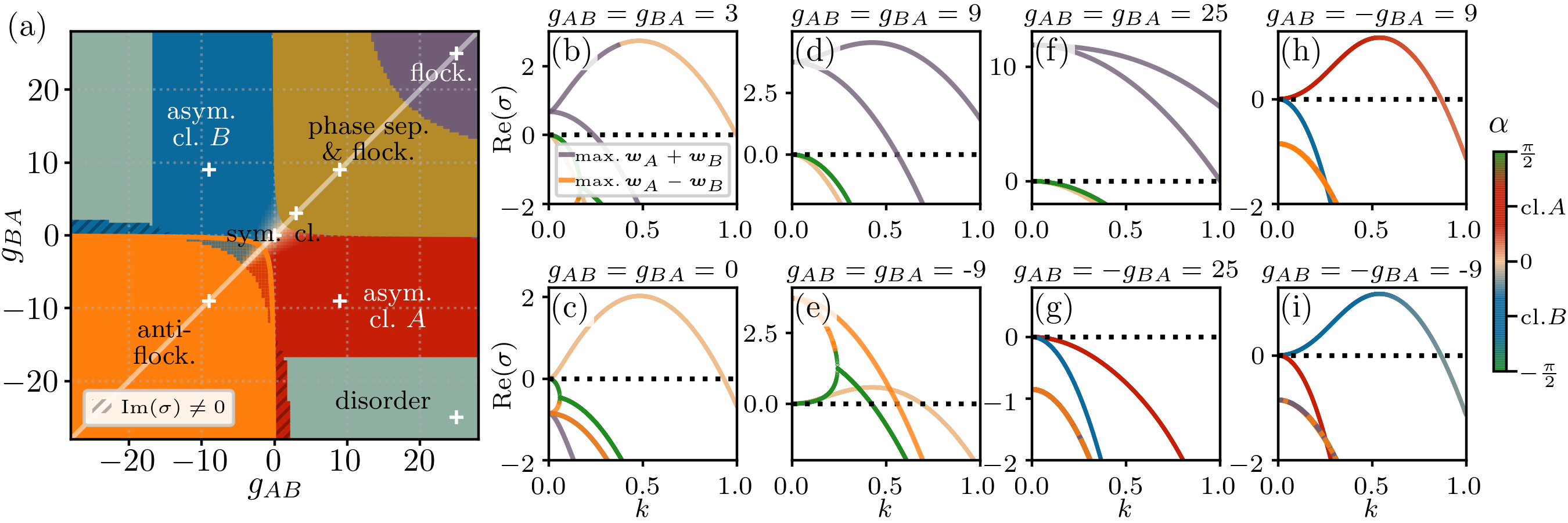}
	\caption{\label{fig:stability_diagram_with_growth_rates}Non-equilibrium phase diagram and respective growth rates. (a) Phase diagram. (b-i) Growth rate for various parameter combinations. The colors indicate the respective eigenvector direction. The phases are determined from linear stability analyses of the disordered base state of the full hydrodynamic Eqs.~(\ref{eq:continuum_eq_density})-(\ref{eq:continuum_eq_polarization}). The white crosses in (a) indicate the parameter combinations whose growth rates are plotted in (b-i). Additional parameters are set to $g_{AA}=g_{BB}=3$, ${\rm Pe}=40$, $z=57.63\,\rho_0^a\,\tau/\ell$, $D_{\rm t}=9$, $D_{\rm r} = 3\cdot 2^{-1/3}$, and $\rho_0^a=4/(5\,\pi)$. See also \cite{kreienkamp_klapp_PRE}.}
\end{figure*}

The characterization is summarized in Table \ref{tab:phases_eigenvalues_eigenvectors_characterization}. Fig.~\ref{fig:stability_diagram_with_growth_rates} shows exemplary real growth rates with indicated largest entries of eigenvectors. Note that (anti-)flocking and (a)symmetric clustering can either occur independent of each other or in combination. For clarity reasons, the phase diagram in Fig.~1 in the main text does not distinguish between these two cases. All instabilities involving flocking instabilities with or without additional clustering instabilities are simply referred to as flocking instabilities.

\begin{table*}
	\begin{tabular}{| p{2.2cm} | p{6.3cm} | p{6cm} |}
		\hline
		non-eq.\ phase & eigenvalues $\sigma_i$ & eigenvector $\bm{v}$ of largest real eigenvalue\\
		\hline
		\hline
		disorder & ${\rm Re}(\sigma_i(k))\leq 0$ for all $k$ and $i=0,...,6$ & --\\
		\hline
		flocking & ${\rm Re}(\sigma_i(k=0))> 0$ for any $i$ & largest entries of eigenvector in $\hat{\bm{w}}^A + \hat{\bm{w}}^B$\\
		\hline
		anti-flocking & ${\rm Re}(\sigma_i(k=0))> 0$ for any $i$ & largest entries of eigenvector in $\hat{\bm{w}}^A - \hat{\bm{w}}^B$\\
		\hline
		sym.\ clustering & ${\rm Re}(\sigma_i(k=0))\leq 0$ for all $i$ and global maximum ${\rm Re}(\sigma_i(k_{\rm max}))$ at $k_{\rm max}>0$ for any $i$ & $\alpha \approx 0$\\
		\hline
		sym.\ demixing & ${\rm Re}(\sigma_i(k=0))\leq 0$ for all $i$ and global maximum ${\rm Re}(\sigma_i(k_{\rm max}))$ at $k_{\rm max}>0$ for any $i$ & $\alpha \approx \pm \pi/2$\\
		\hline
		asym.\ cl.\ $A$ & ${\rm Re}(\sigma_i(k=0))\leq 0$ for all $i$ and global maximum ${\rm Re}(\sigma_i(k_{\rm max}))$ at $k_{\rm max}>0$ for any $i$ & $0<\alpha < \pi/2$\\
		\hline
		asym.\ cl.\ $B$ & ${\rm Re}(\sigma_i(k=0))\leq 0$ for all $i$ and global maximum ${\rm Re}(\sigma_i(k_{\rm max}))$ at $k_{\rm max}>0$ for any $i$ & $-\pi/2<\alpha < 0$\\
		\hline
	\end{tabular}
	\caption{\label{tab:phases_eigenvalues_eigenvectors_characterization}Characterization of non-equilibrium phases in terms of eigenvalues and eigenvector corresponding to largest eigenvalue. The angle $\alpha=\arccos(\bm{v}_{\rho} \cdot \bm{x}_{\rm con})$ with $\bm{v}_{\rho}=(\hat{\rho}^A+\hat{\rho}^B, \, \hat{\rho}^A-\hat{\rho}^B)^{\rm T}$ and $\bm{x}_{\rm con}=(1,0)^{\rm T}$ indicates the type of phase separation. See also \cite{kreienkamp_klapp_PRE}.}
\end{table*}  

In our system with relatively weak intraspecies alignment couplings of $g_{AA}=g_{BB}=3$, the eigenvalues are real for a majority of intraspecies coupling strengths. For eigenvalues with positive real part and non-zero imaginary part, instabilities are oscillatory, indicating non-stationary emerging phases.

\subsection{Flocking behavior}
\label{ssec:mean-field_flocking_behavior}
Flocking instabilities are related to unstable long-wavelength ($k=0$) fluctuations of polarizations. The resulting mean-field phase diagram at $k=0$ is shown in Fig.~1(a) in the main text. For the corresponding hydrodynamic description we ignore all gradient terms in Eqs.~(\ref{eq:continuum_eq_density})-(\ref{eq:continuum_eq_polarization}). This yields
\begin{equation}
	\begin{split}
		\partial_t \rho^a &= 0 \quad \text{with} \quad \rho^a = 1\\
		\partial_t \bm{w}^a &= -D_{\rm r} \, \bm{w}^a + \sum_b g'_{ab} \, \rho^a \, \bm{w}^b - \sum_{b,c} \frac{g'_{ab}\,g'_{ac}}{2\,D_{\rm r}} \, \bm{w}^a \, (\bm{w}^b \cdot \bm{w}^c) .
	\end{split}
\end{equation}
The densities are conserved with $\rho^A = \rho^B = 1$. At $k=0$, their perturbations can therefore be ignored. The polarization dynamics can be written in terms of the matrix equation 
\begin{equation}
	\label{eq:mean-field_polarization_matrix_equation}
	\partial_t \begin{pmatrix}
		\bm{w}^A \\
		\bm{w}^B
	\end{pmatrix} = \begin{pmatrix}
		-D_{\rm r} + g'_{AA} - \frac{1}{2\,D_{\rm r}}\, \bm{\mathcal{Q}}_A^2 & g'_{AB} \\
		g'_{BA} & -D_{\rm r} + g'_{BB} - \frac{1}{2\,D_{\rm r}}\, \bm{\mathcal{Q}}_B^2
	\end{pmatrix} \cdot \begin{pmatrix}
		\bm{w}^A \\
		\bm{w}^B
	\end{pmatrix}
\end{equation}
with
\begin{equation}
	\bm{\mathcal{Q}}_A = g'_{AA}\,\bm{w}^A + g'_{AB}\,\bm{w}^B \qquad \text{and} \qquad \bm{\mathcal{Q}}_B = g'_{BB}\,\bm{w}^B + g'_{BA}\,\bm{w}^A .
\end{equation}
Instead of the 2d-vector representation, $\bm{w}^a = w_0^a \, (\cos(\theta^a), \sin(\theta^a))^{\rm T}$, we now represent the polarization field as a complex number $\bm{w}^a = w_0^a \, {\rm e}^{i\,\theta^a}$ with amplitude $w_0^a = \sqrt{(w_x^a)^2+ (w_y^a)^2}$ and phase $\theta^a=\arctan(w_y^a/w_x^a)$. Separating real and complex parts, we get the evolution equations for the amplitude,
\begin{equation}
	\begin{split}
		\partial_t w_0^A = & -D_{\rm r} \, w_0^A + g'_{AA} \,  w_0^A + g'_{AB} \, w_0^B \, \cos(\theta^B-\theta^A) - \frac{g'_{AA}\,g'_{AA}}{2\,D_{\rm r}} \, (w_0^A)^3 \\
		&  - \frac{g'_{AA}\,g'_{AB}}{D_{\rm r}} \, (w_0^A)^2\,w_0^B \,  \cos(\theta^B-\theta^A) - \frac{g'_{AB}\,g'_{AB}}{2\,D_{\rm r}} \, w_0^A\, (w_0^B)^2,
	\end{split}
\end{equation}
and phase,
\begin{equation}
	\partial_t \theta^A = \frac{w_0^B \, \sin(\theta^B - \theta^A) \, g'_{AB}}{w_0^A} \, \left(\rho^A - \frac{g'_{AA}}{D_{\rm r}} (w_0^A)^2 \right) .
\end{equation}
The equations for $w_0^B$ and $\theta^B$ are obtained equivalently.

Generally, fixed points of these equations are characterized by $\partial_t w_0^a=0$ and $\partial_t \theta^a=0$. The disordered fixed point corresponds to $w_0^A=w_0^B = 0$ with arbitrary $\theta^A,\theta^B$. For $w_0^a \neq 0$, fixed points are given for $\theta^A=\theta^B$. These either correspond to (parallel) flocking if $w_0^A \, w_0^B > 0$ or to (anti-parallel) anti-flocking if $w_0^A \, w_0^B < 0$.

We focus on the disordered fixed point, which yields the flocking transition. Here, we can focus on the amplitude equations alone, such that we end up with a $2\times 2$-problem. We look at the linear stability of amplitude perturbations $w_0^a = 0 + \epsilon\,\tilde{w}^a$. The corresponding eigenvalue equation for $\tilde{w}^a = \tilde{w}^a_0 \, {\rm e}^{\sigma\,t}$ is
\begin{equation}
	\label{eq:mean-field_polarization_eigenvalue_equation}
	\sigma \, \begin{pmatrix}
		\tilde{w}_0^A \\
		\tilde{w}_0^B
	\end{pmatrix} = \begin{pmatrix}
		-D_{\rm r} + g'_{AA}  & g'_{AB} \\
		g'_{BA} & -D_{\rm r} + g'_{BB}
	\end{pmatrix} \cdot \begin{pmatrix}
		\tilde{w}_0^A \\
		\tilde{w}_0^B
	\end{pmatrix}.
\end{equation}
The eigenvalues
\begin{equation}
	\begin{split}
		\sigma_{1/2} =  \frac{1}{2} \Bigg( & (-2\,D_{\rm r} + g'_{AA} + g'_{BB}) \pm \sqrt{\left((g'_{AA} - D_{\rm r}) - (g'_{BB} - D_{\rm r}) \right)^2 + 4\, g'_{AB} \, g'_{BA}} \Bigg)
	\end{split}
\end{equation}
mark the flocking transition lines. The type of flocking is determined by the eigenvector $\bm{v}^{\rm fl} = (v^A, v^B)^{\rm T}$, which corresponds to the largest positive eigenvalue. If the product $v^A\,v^B>0$, the $k-0$-instability leads to flocking, where $\bm{P}_A || \bm{P}_B$. For $v^A\,v^B<0$, anti-flocking emerges with $\bm{P}_A || -\bm{P}_B$. Importantly, on the continuum level the (anti-)flocking instabilities do not depend on repulsion effects ($\sim z$). We also note that the linear stability analysis around homogeneous phases without steric repulsion cannot explain the emergence of polarized high-density bands observed in Vicsek-like systems \cite{vicsek_1995_novel_type_phase_transition,solon_tailleur_2015_phase_separation_flocking_models}.

\subsection{Coarse-grained density dynamics}
The linear stability analysis of the full six-dimensional dynamics in section \ref{ssec:linear_stability_analysis_non-equilibirum_phase_characterization} shows that asymmetric density instabilities are mainly present outside the flocking regime, see Fig.~\ref{fig:stability_diagram_with_growth_rates}. To explain the clustering behavior, we consider a simplified coarse-grained equation of the dynamics in terms of density fields alone under adiabatical approximation of polarization fields.

The starting point are the full hydrodynamic Eqs.~\eqref{eq:continuum_eq_density}-\eqref{eq:continuum_eq_polarization} for $\rho^a$ and $\bm{w}^a$. Outside the flocking regime, no orientational order emerges. On large length and time scales, we can therefore (adiabatically) eliminate temporal and spatial derivatives as well as higher-order moments of the polarization densities $\bm{w}^a$.
Under these approximations, Eq.~\eqref{eq:continuum_eq_polarization} for the polarization densities $\bm{w}^a_{\rm ad}$ simplifies to
\begin{equation}
	0 = - \frac{1}{2} \, \nabla \,\big(v^{\rm eff}(\rho)\, \rho^a\big)  - D_{\rm r }\, \bm{w}_{\rm ad}^{a} + \sum_b g'_{ab} \, \rho^a\, \bm{w}_{\rm ad}^b 
\end{equation}
with $v^{\rm eff}(\rho)={\rm Pe}- z\,\rho$ and $\rho=\rho^A + \rho^B$. 
For $a=A$, this yields 
\begin{equation}
	\label{eq:coarse-graining_wA_depends_on_wB}
	\begin{split}
		\bm{w}_{\rm ad}^A &= \frac{1}{g'_{AA}\,\rho^A - D_{\rm r}} \left(\frac{1}{2} \nabla \,\big(v^{\rm eff}(\rho)\, \rho^A\big) - g'_{AB}\,\rho^A\,\bm{w}_{\rm ad}^B  \right).
	\end{split}
\end{equation}
Inserting the respective equation for $\bm{w}_{\rm ad}^B$ yields
\begin{equation}
	\bm{w}_{\rm ad}^A(\rho^A, \rho^B) = \frac{-v^{\rm eff}(\rho) \big((D_{\rm r} - g'_{BB}\,\rho^B) \nabla \rho^A + g'_{AB}\,\rho^A \, \nabla \rho^B \big) + \rho^A \, (D_{\rm r} + (g'_{AB} - g'_{BB})\, \rho^B) \, z\, \nabla \rho}{2\, \big(D_{\rm r}^2 + (-g'_{AB}\,g'_{BA} + g'_{AA}\, g'_{BB})\,\rho^A\,\rho^B - D_{\rm r}\,(g'_{AA}\,\rho^A + g'_{BB}\,\rho^B) \big)}
\end{equation}
and 
\begin{equation}
		\bm{w}_{\rm ad}^B(\rho^A, \rho^B) = \frac{-v^{\rm eff}(\rho) \big((D_{\rm r} - g'_{AA}\,\rho^A) \nabla \rho^B + g'_{BA}\,\rho^B \, \nabla \rho^A \big) + \rho^B \, (D_{\rm r} + (g'_{BA} - g'_{AA})\, \rho^A) \, z\, \nabla \rho}{2\, \big(D_{\rm r}^2 + (-g'_{AB}\,g'_{BA} + g'_{AA}\, g'_{BB})\,\rho^A\,\rho^B - D_{\rm r}\,(g'_{AA}\,\rho^A + g'_{BB}\,\rho^B) \big)},
\end{equation}
equivalently.
These equations can now be inserted into the continuity Eq.~\eqref{eq:continuum_eq_density} for $\rho^A$ and $\rho^B$. For $\rho^A \pm \rho^B$, the equations read
\begin{equation}
	\label{eq:coarse-grained_density_equation}
	\underbrace{\partial_t (\rho^A \pm \rho^B)}_{=(1)} = \underbrace{-\nabla \cdot (v^{\rm eff}(\rho^A + \rho^B) \, (\bm{w}_{\rm ad}^A \pm \bm{w}_{\rm ad}^B))}_{=(2)} + \underbrace{D_{\rm t}\, \nabla^2 (\rho^A \pm \rho^B)}_{=(3)} .
\end{equation}
with
\begin{equation}
	\bm{w}_{\rm ad}^A(\rho^A, \rho^B) \pm \bm{w}_{\rm ad}^B(\rho^A, \rho^B) = \frac{\mathcal{N}_{\pm}(\rho^A, \rho^B)}{2\, \big(D_{\rm r}^2 + (-g'_{AB}\,g'_{BA} + g'_{AA}\, g'_{BB})\,\rho^A\,\rho^B - D_{\rm r}\,(g'_{AA}\,\rho^A + g'_{BB}\,\rho^B) \big)},
\end{equation}
where
\begin{equation}
	\begin{split}
		\mathcal{N}_{\pm}(\rho^A, \rho^B) =&  -v^{\rm eff}(\rho) \, \big[ \big(D_{\rm r} + (\pm g'_{BA} - g'_{BB})\,\rho^B\big) \nabla \rho^A \pm \big(D_{\rm r} + ( \pm g'_{AB}- g'_{AA})\,\rho^A\big) \nabla \rho^B \big] \\
		& + \big[D_{\rm r}\,(\rho^A \pm \rho^B) + (\mp g'_{AA} - g'_{BB} + g'_{AB} \pm g'_{BA})\,\rho^A \, \rho^B\big] \, z\, \nabla \rho.
	\end{split}
\end{equation}

\subsubsection{Linear stability matrix}
For the linear stability analysis around the homogeneous disordered phase, i.e., $\rho^a = 1 + \epsilon\,\rho'^a$, we look at the terms (1)-(3) in Eq.~\eqref{eq:coarse-grained_density_equation} individually. (1) and (3) are straightforward and can be easily transformed into the relevant quantities in Fourier space. Term (2) needs more modification. We have 
\begin{equation}
	\begin{split}
		(2): \quad -\nabla& \cdot (v^{\rm eff}(\rho^A + \rho^B) \, (\bm{w}_{\rm ad}^A \pm \bm{w}_{\rm ad}^B)) = -(\bm{w}_{\rm ad}^A \pm \bm{w}_{\rm ad}^B) \cdot \nabla v^{\rm eff}(\rho^A + \rho^B) - v^{\rm eff}(\rho^A + \rho^B) \, \nabla \cdot (\bm{w}_{\rm ad}^A \pm \bm{w}_{\rm ad}^B) .
	\end{split}
\end{equation}
We now insert $\rho^a = 1 + \epsilon\,\rho'^a$ into the equations for $\bm{w}_{\rm ad}^A \pm \bm{w}_{\rm ad}^B$. We then perform a Taylor expansion of $\bm{w}_{\rm ad}^A \pm \bm{w}_{\rm ad}^B$ around small $\epsilon$. The result is
\begin{equation}
	\bm{w}_{\rm ad}^A \pm \bm{w}_{\rm ad}^B = \underbrace{\bm{w}_{{\rm ad},0}^A \pm \bm{w}_{{\rm ad},0}^B}_{= \bm{w}_{\rm ad}^A(1,1) \pm \bm{w}_{\rm ad}^B(1,1) = 0} +\  \bm{f}_1^{\pm}(\rho'^A, \rho'^B) \, \epsilon + O(\epsilon^2) 
\end{equation}
with
\begin{equation}
	\bm{f}_1^{\pm}(\rho'^A, \rho'^B) = \gamma^A_{\pm} \, \nabla \rho'^A + \gamma^B_{\pm} \, \nabla \rho'^B.
\end{equation}
The prefactors are 
\begin{equation}
		\gamma^A_{+} = \frac{D_{\rm r} \, \big(4\,z - {\rm Pe}\big)+ \big[(g'_{AB}-g'_{AA})\, z + g'_{BA} \, (3 \, z -{\rm Pe}) + g'_{BB} \, ({\rm Pe} - 3\, z)\big]}{\mathcal{D}_0},
\end{equation}
\begin{equation}
		\gamma^B_{+} = \frac{D_{\rm r} \, \big(4\,z-{\rm Pe}\big)+ \big[(g'_{BA}-g'_{BB})\, z + g'_{AB} \, (3 \, z -{\rm Pe}) + g'_{AA} \, ({\rm Pe} - 3\, z)\big]}{\mathcal{D}_0},
\end{equation}
\begin{equation}
		\gamma^A_{-} = \frac{-D_{\rm r} \, \big({\rm Pe}-2\,z\big)+ \big[(g'_{AB}+g'_{AA})\, z + g'_{BA} \, ({\rm Pe}-3 \, z) + g'_{BB} \, ({\rm Pe} - 3\, z)\big]}{\mathcal{D}_0},
\end{equation}
and
\begin{equation}
		\gamma^B_{-} = \frac{D_{\rm r} \, \big({\rm Pe}-2\,z\big)+ \big[-(g'_{BA}+g'_{BB})\, z + g'_{AB}\, (-{\rm Pe}+3 \, z) + g'_{AA} \, (-{\rm Pe} + 3\, z)\big]}{\mathcal{D}_0},
\end{equation}
where
\begin{equation}
	\mathcal{D}_0 = 2\, \big(D_{\rm r}^2 + (-g'_{AB}\,g'_{BA} + g'_{AA}\, g'_{BB}) - D_{\rm r}\,(g'_{AA} + g'_{BB}) \big) .
\end{equation}
For term $(2)$ in Eq.~\eqref{eq:coarse-grained_density_equation} we thus have
\begin{equation}
	-\nabla \cdot (v^{\rm eff}(\rho^A + \rho^B) \, (\bm{w}_{\rm ad}^A \pm \bm{w}_{\rm ad}^B)) = \epsilon \, \Big(-V \, \nabla \cdot \bm{f}_1^{\pm}(\rho'^A, \rho'^B) \Big) + O(\epsilon^2),
\end{equation}
where $V = {\rm Pe} - 2\,z$.
In Fourier space, the time-evolution of the density equations read
\begin{equation}
	\sigma \, (\hat{\rho}_A \pm \hat{\rho}_B) = V \, \big(\gamma_{\pm}^A \, \hat{\rho}^A + \gamma_{\pm}^B \, \hat{\rho}^B \big)\, k^2 - D_{\rm t}\,(\hat{\rho}_A \pm \hat{\rho}_B)\,k^2 .
\end{equation}
This can be rewritten in terms of the eigenvalue equation
\begin{equation}
	\label{eq:coarse-grained_dynamics_eigenvalue_equation}
	\sigma \begin{pmatrix}
		\hat{\rho}_A + \hat{\rho}_B \\
		\hat{\rho}_A - \hat{\rho}_B
	\end{pmatrix} = \begin{pmatrix}
		\big(\frac{V}{2}\, (\gamma_+^A + \gamma_+^B) - D_{\rm t}\big)\,k^2 & \frac{V}{2}\, (\gamma_+^A - \gamma_+^B)\,k^2 \\
		\frac{V}{2}\, (\gamma_-^A + \gamma_-^B)\,k^2 & \big(\frac{V}{2}\, (\gamma_-^A - \gamma_-^B) - D_{\rm t}\big)\,k^2
	\end{pmatrix} \cdot \begin{pmatrix}
		\hat{\rho}_A + \hat{\rho}_B \\
		\hat{\rho}_A - \hat{\rho}_B
	\end{pmatrix}.
\end{equation}
For $g'_{AA}=g'_{BB}=g$ and $D_{\rm t}=0$, the matrix can be simplified to those given in the main text for reciprocal ($g'_{AB}=g'_{BA}=\kappa$) and fully anti-symmetric ($g'_{AB}=-g'_{BA}=\delta$) couplings.

\begin{figure*}
	\includegraphics[width=\linewidth]{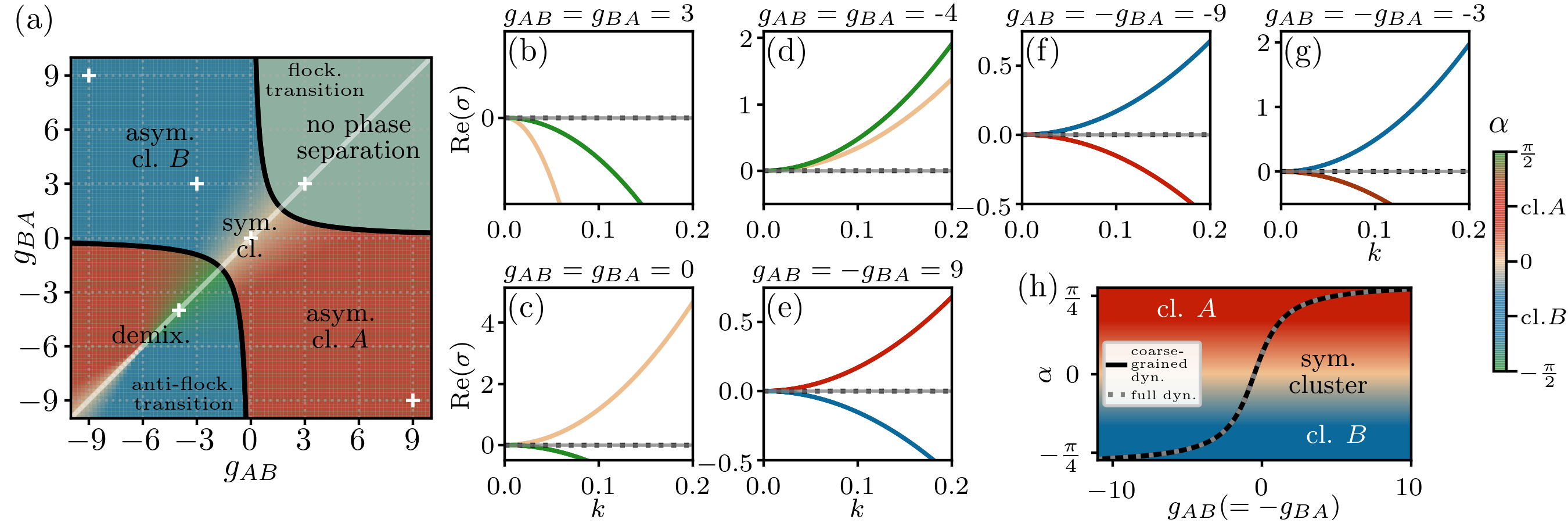}
	\caption{\label{fig:coarse_grained_density_stability_diagram_with_growth_rates}Non-equilibrium phase diagram and respective growth rates of the coarse-grained density evolution with adiabatic approximation of polarization [Eq.~(\ref{eq:coarse-grained_density_equation})]. (a) Phase diagram. (b-g) Growth rate for various parameter combinations. The colors indicate the respective eigenvector direction. (h) The clustering angle $\alpha$ for different $g_{AB}$ in the full continuum description [Eqs.~(\ref{eq:continuum_eq_density})-(\ref{eq:continuum_eq_polarization})] and coarse-grained density description [Eq.~(\ref{eq:coarse-grained_density_equation})]. The white crosses in (a) indicate the parameter combinations whose growth rates are plotted in (b-g). Additional parameters are set to $g_{AA}=g_{BB}=3$, ${\rm Pe}=40$, $z=57.63\,\rho_0^a\,\tau/\ell$, $D_{\rm t}=0$, $D_{\rm r} = 3\cdot 2^{-1/3}$, and $\rho_0^a=4/(5\,\pi)$.}
\end{figure*}

\subsubsection{Linear stability results}
To determine and classify the density instabilities, we now focus on eigenvalues and -vectors at small wave number $k$. Being only interested in phase separation phenomena, we basically follow the same clustering classification as summarized in Table \ref{tab:phases_eigenvalues_eigenvectors_characterization}: A density instability is indicated by ${\rm Re}(\sigma_{\rho})>0$ at $k>0$. The corresponding eigenvectors $\bm{v} = (\hat{\rho}_A + \hat{\rho}_B, \hat{\rho}_A - \hat{\rho}_B)^{\rm T}$ indicate the type of phase separation. The direction of the eigenvector corresponding to the largest eigenvalue is given by the clustering angle $\alpha = \arccos(\bm{v}\cdot (1,0)^{\rm T})$. Symmetric clustering is indicated by $\alpha = 0$ and full symmetric demixing by $\alpha=\pm\pi/2$. Asymmetric clustering (i.e., partial demixing) of predominantly species $A$ ($B$) corresponds to $0<\alpha<\pi/2$ ($-\pi/2 < \alpha < 0$).

The resulting non-equilibrium phase diagram in the coarse-grained density description for arbitrary combinations $g_{AB},g_{BA}$ is shown in Fig.~\ref{fig:coarse_grained_density_stability_diagram_with_growth_rates}(a). Exemplary eigenvalues and largest eigenvector components are plotted in Fig.~\ref{fig:coarse_grained_density_stability_diagram_with_growth_rates}(b)-(g).

Outside the (anti-)flocking regimes, reciprocal couplings lead to symmetric clustering. Moving into non-reciprocal regime but staying outside of flocking regimes, asymmetric clustering of species $A$ (for $g_{AB}>0, g_{BA}<0$) or species $B$ (for $g_{AB}<0, g_{BA}>0$) is predicted. As seen in Fig.~\ref{fig:coarse_grained_density_stability_diagram_with_growth_rates}(h), the clustering angle of the coarse-grained density description perfectly matches the one of the full continuum description [Eqs.~(\ref{eq:continuum_eq_density})-(\ref{eq:continuum_eq_polarization})].

Inside the flocking regime, no density instabilities are predicted. Inside the anti-flocking regime, demixing is predicted for reciprocal couplings and asymmetric clustering for non-reciprocal couplings. Here, the asymmetric clustering is predicted with ``opposite'' clustering preference than outside the anti-flocking regime. These results do not correspond to those obtained from the full dynamics [Eqs.~(\ref{eq:continuum_eq_density})-(\ref{eq:continuum_eq_polarization})]. This is not unexpected, as the assumption of small-magnitude polarization fields does not hold when flocking instabilities are present.

Thus, outside the flocking regime, the coarse-grained density dynamics can predict the correct phase separation behavior. However, inside the flocking regime, the coarse-grained density dynamics does not yield the phase behavior expected from particle simulations. Instead, the full six-dimensional problem must be considered to determine a correct combination of flocking and density instabilities without incorrect classifications. In particular, the combination of flocking and phase separation for large $g_{AB},g_{BA}$ is not captured by the reduced description.

\subsection{Relation to exceptional points}
Exceptional transitions have been related to parity-time symmetry breaking transitions in non-reciprocal scalar \cite{You_Baskaran_Marchetti_2020_pnas,saha_scalar_active_mixtures_2020} and strongly coupled polar active systems \cite{fruchart_2021_non-reciprocal_phase_transitions}. In conserved scalar systems, exceptional points mark a transition from static patterns to traveling waves \cite{You_Baskaran_Marchetti_2020_pnas}. In polar systems, exceptional transitions separate regimes of (anti-)parallel (anti-)flocking in constant direction of polarization and chiral phases, where the polarization direction rotates in time \cite{fruchart_2021_non-reciprocal_phase_transitions}. Here, we briefly discuss the appearance of exceptional points in the present system.

\begin{figure}
	\includegraphics[width=1\linewidth]{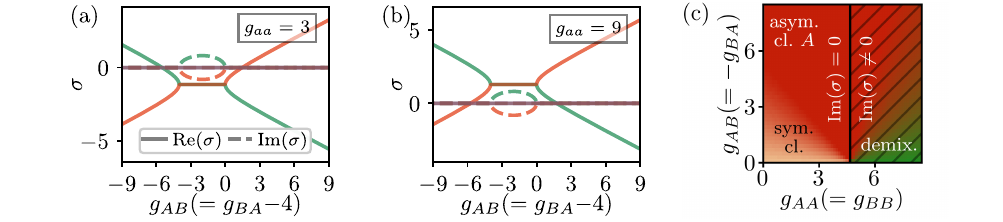}
	\caption{\label{fig:eigenvalues_relation_to_exceptional_points}Eigenvalues and phase diagram at wave number $k=0$ for different intraspecies coupling strengths $g_{aa}$. The eigenvalues of the isotropic disordered base state are shown for $g_{AB}=g_{BA}-4$ and (a) $g_{AA}=g_{BB}=3$,  (b) $g_{AA}=g_{BB}=9$. At $g_{AB}=0$ and $g_{AB}=-4$, four eigenvalues coalesce and form two complex conjugate pairs. The eigenvectors corresponding to the complex conjugate pairs become parallel. (a) Below the flocking transition line, the real part of the eigenvalues, ${\rm Re}(\sigma)$, is negative as long as ${\rm Im}(\sigma)\neq 0$. (b) For stronger intraspecies couplings, the formation of complex conjugate pairs with parallel eigenvectors indicates instabilities with exceptional points since ${\rm Re}(\sigma)>0$ as long as ${\rm Im}(\sigma)\neq 0$. (c) The phase diagram for $g_{AB}=-g_{BA}$ shows that below the flocking transition line at $g_{aa}\approx 4.7$, all positive eigenvalues are real and no $k=0$-instabilities exist. Above the transition line, oscillatory $k=0$-instabilities appear. The color indicates the degree of (a)symmetric clustering.}
\end{figure}

By definition, exceptional points are points, where eigenvalues of the linear stability matrix coalesce and the eigenvectors become parallel \cite{el-ganainy_christodoulides_2018_non-hermitian_physics_PT-symmetry}. We consider the isotropic disordered state (with constant density and vanishing polarization) as fixed point. The linear stability of the fixed point is determined by matrix \eqref{eq:stability_matrix}. In Fig.~\ref{fig:eigenvalues_relation_to_exceptional_points}, we show the corresponding six eigenvalues $\sigma$ at $k=0$ as a function of $g_{AB}$. We choose $g_{BA}=g_{AB}-d$ with $d=4$. Further, we set $g_{AA}=g_{BB}=g_{aa}=3$ in Fig.~\ref{fig:eigenvalues_relation_to_exceptional_points}(a) and $g_{aa}=9$ in Fig.~\ref{fig:eigenvalues_relation_to_exceptional_points}(b). The two density-related eigenvalues are zero at $k=0$, regardless of $g_{AB}$ (compare Sec.~\ref{ssec:linear_stability_analysis_non-equilibirum_phase_characterization}). The other four eigenvalues come in pairs for the here considered cases (and overlap in Fig.~\ref{fig:eigenvalues_relation_to_exceptional_points}).

Regardless of $g_{aa}$, we can make the following observation. For $g_{AB}<-d=-4$ and $g_{AB}>0$, the two pairs of eigenvalues are distinct and real (with vanishing imaginary part). At $g_{AB}=0$ and $g_{AB}=-4$, the four (non-zero) eigenvalues coalesce and form two complex conjugate pairs. Hence, for $-4<g_{AB}<0$, all non-zero eigenvalues have the same real parts and non-zero imaginary parts. The points of eigenvalue coalescence are $g_{AB}=0,-4$. The eigenvectors corresponding to the complex conjugate pairs become parallel at these points. 

Depending on $g_{aa}$, the points of eigenvalue coalescence have different implications. Below the flocking transition line [``weak intraspecies couplings'' with $g_{aa}=3$, Fig.~\ref{fig:eigenvalues_relation_to_exceptional_points}(a)], the real part of the eigenvalues is negative as long as ${\rm Im}(\sigma)\neq 0$. For stronger intraspecies couplings [$g_{aa}=9$, Fig.~\ref{fig:eigenvalues_relation_to_exceptional_points}(b)], ${\rm Re}(\sigma)>0$ as long as ${\rm Im}(\sigma)\neq 0$. The formation of complex conjugate pairs with parallel eigenvectors then indicates instabilities with exceptional points. Hence, exceptional transitions related to instabilities only appear for strong intraspecies couplings. In particular, for our system and chosen parameters (see Sec.~\ref{sapp:parameter_choice_continuum}), the flocking transition line $g_{aa}\approx 4.7$ marks the strength of when exceptional transitions play a role in the stability of the disordered base state. 

Fig.~\ref{fig:eigenvalues_relation_to_exceptional_points}(c) shows the corresponding stability diagram in the $g_{aa} - g_{AB}$-plane for the fully antisymmetric case $g_{AB} = -g_{BA}$. 
Below the flocking transition line ($g_{aa}\lesssim 4.7$), no $k=0$-instabilities are predicted for the here considered case of $g_{AB}\,g_{BA}<0$.
Beyond the critical flocking line ($g_{aa}\gtrsim 4.7$), eigenvalues have positive real parts at $k=0$. This marks the exceptional transition. These $k=0$-instabilities are oscillatory (with non-zero imaginary part) for $g_{AB}\,g_{BA}<0$. Asymmetric clustering is predicted below and above the flocking transition. Yet, towards the reciprocal limit, symmetric clustering below the flocking transition changes to demixing above the flocking transition.

Additionally to the here considered disordered base state, one can also study the linear stability and exceptional transitions of homogeneous \mbox{(anti-)}flocking states, see \cite{fruchart_2021_non-reciprocal_phase_transitions}. Such homogeneous \mbox{(anti-)}flocking base states only exist in regimes, where $k=0$-instabilities occur. The effect of non-reciprocal alignment on density dynamics in the presence of exceptional points at larger intraspecies couplings strengths will be considered in our future research.

\section{\label{app:impact_parameters}Impact of parameters}
The active repulsive mixture with non-reciprocal alignment has various control parameters. We here discuss their impact on the collective behavior based on the linear stability analysis of the homogeneous disordered base state. The role of repulsive interactions on a particle-level are discussed in \cite{kreienkamp_klapp_PRE}.

\subsection{Role of orientational parameters}
As seen in Sec.~\ref{ssec:mean-field_flocking_behavior}, alignment strengths and rotational diffusion determine the flocking behavior. For large alignment strengths, a $k=0$-instability is predicted. This flocking instability indicates the emergence of non-zero global polarization. Rotational diffusion counteracts flocking.

\subsection{Role of translational parameters}
Translational diffusion damps perturbations at large wave numbers. Increasing translational diffusion decreases the regimes of density-instabilities in the $g_{AB}-g_{BA}$-plane. Yet, as long as instabilities are predicted at all, translational diffusion has no impact on the type of instability. This is because we determine the type of phase separation in terms of the clustering angle at small $k$.

\begin{figure}
	\includegraphics[width=1\linewidth]{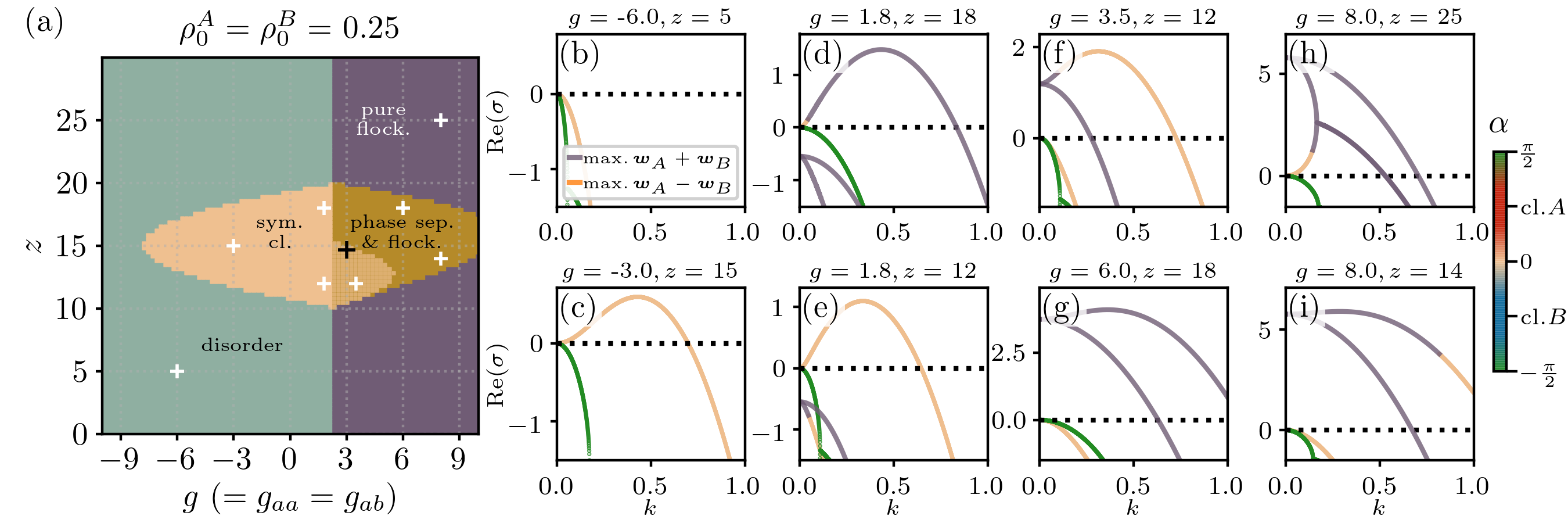}
	\caption{\label{fig:impact_parameter_z_one-species_system}Non-equilibrium phase diagram and respective growth rates of the effective-one-species system at different coupling strengths $g=g_{AA}=g_{BB}=g_{AB}=g_{BA}$ and velocity-reduction parameters $z$. (a) Phase diagram. (b-i) Growth rates for various parameter combinations. The colors indicate the respective eigenvector direction. The white crosses in (a) indicate the parameter combinations whose growth rates are plotted in (b-i). The black cross in (a) marks the parameters chosen in the effective one-species system considered in this study ($g=3$, $z=57.63$) The phase diagram is obtained from the full continuum description [Eqs.~(\ref{eq:continuum_eq_density})-(\ref{eq:continuum_eq_polarization})]. Additional parameters are set to ${\rm Pe}=40$, $D_{\rm t}=9$, $D_{\rm r} = 3\cdot 2^{-1/3}$, and $\rho_0^a=4/(5\,\pi)$.}
\end{figure}

The self-propulsion speed of particles is given by $v^{\rm eff}(\rho^A + \rho^B) = {\rm Pe} - z\,(\rho^A + \rho^B)$. Hence, the velocity-reduction parameter $z$ determines how much the effective self-propulsion speed of particles is reduced in crowded environments. The impact of $z$ is shown in Fig.~\ref{fig:impact_parameter_z_one-species_system} for a (reciprocal) effective one-species system with $\rho_0^A = \rho_0^B$ for different $g=g_{AA}=g_{BB}=g_{AB}=g_{BA}$. Flocking only emerges for sufficiently large alignment $g$ -- independent of $z$. However, $z$ determines whether density instabilities are present or not. For a range of intermediate $z$-values, clustering appears either without or in combination with flocking instabilities. If $z$ is too small or too large, no clustering is predicted. This is in agreement with continuum studies \cite{speck_2015_dynamical_mean_field_phase_separation,elena_2021_phase_separation_self-propelled_disks,kreienkamp_klapp_2022_clustering_flocking_chiral_active_particles_non-reciprocal_couplings}. The parameters chosen in the main text are indicated by the black cross in Fig.~\ref{fig:impact_parameter_z_one-species_system} and place us in the middle of the density-instability region.

\begin{figure}
	\includegraphics[width=1\linewidth]{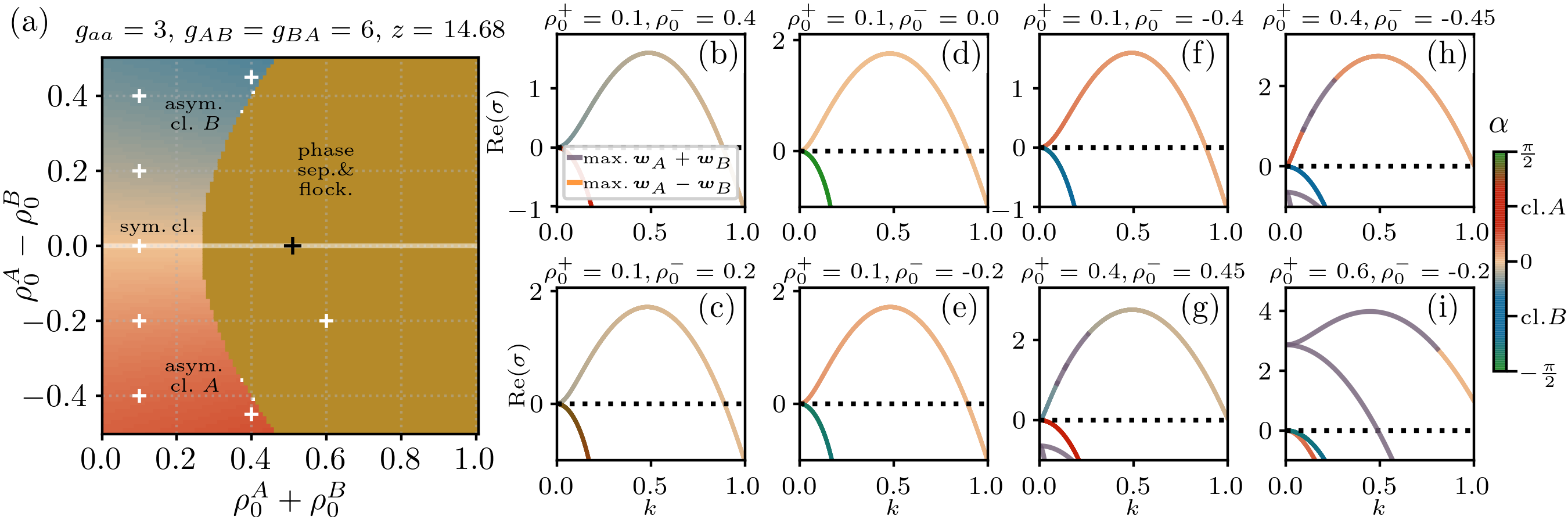}
	\caption{\label{fig:impact_parameter_rho_reciprocal}Non-equilibrium phase diagram and respective growth rates of the reciprocal system at different compositions of species $A$ and $B$. The coupling strengths are set to $g_{AA}=g_{BB} = 3$ and $g_{AB}=g_{BA} = 6$. (a) Phase diagram. (b-i) Growth rates for various compositions. The colors indicate the respective eigenvector direction. The white crosses in (a) indicate the parameter combinations whose growth rates are plotted in (b-i). The black cross in (a) marks the composition chosen in the main text ($\rho_0^A = \rho_0^B = 0.4/\pi$) The phase diagram is obtained from the full continuum description [Eqs.~(\ref{eq:continuum_eq_density})-(\ref{eq:continuum_eq_polarization})]. Additional parameters are set to ${\rm Pe}=40$, $z=14.68$, $D_{\rm t}=9$.}
\end{figure}

\subsection{Role of density composition}
\begin{figure}
	\includegraphics[width=1\linewidth]{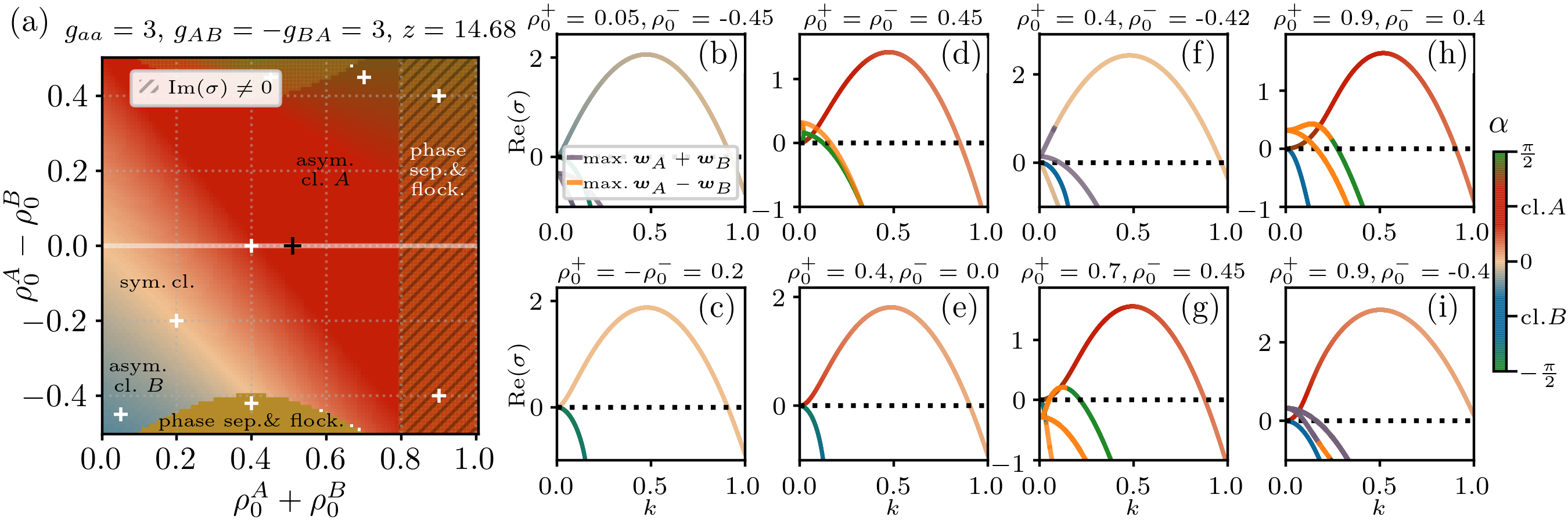}
	\caption{\label{fig:impact_parameter_rho_non-reciprocal}Non-equilibrium phase diagram and respective growth rates of the non-reciprocal system at different compositions of species $A$ and $B$. Intraspecies couplings are $g_{aa}=3$, interspecies couplings are anti-symmetric with $g_{AB}=-g_{BA} = 3$. (a) Phase diagram. (b-i) Growth rates for various compositions. The colors indicate the respective eigenvector direction. The white crosses in (a) indicate the parameter combinations whose growth rates are plotted in (b-i). The black cross in (a) marks the composition chosen in the main text ($\rho_0^A = \rho_0^B = 0.4/\pi$) The phase diagram is obtained from the full continuum description [Eqs.~(\ref{eq:continuum_eq_density})-(\ref{eq:continuum_eq_polarization})]. Additional parameters are set to ${\rm Pe}=40$, $z=14.68$, $D_{\rm t}=9$.}
\end{figure}

We now consider variations in the total density and composition of species $A$ and $B$. In an effective one-component system with $g_{AA}=g_{BB}=g_{AB}=g_{BA}$, the ratio of particle species does not play any role. However, the density composition significantly affects the dynamics as soon as the species differ from each other -- both in reciprocal systems with $g_{AA}=g_{BB}\neq g_{AB}=g_{BA}$ and non-reciprocal systems with $g_{AB}\neq g_{BA}$. The phase diagram for the reciprocal system is shown in Fig.~\ref{fig:impact_parameter_rho_reciprocal} and for the non-reciprocal one in Fig.~\ref{fig:impact_parameter_rho_non-reciprocal}.

Regardless of the (non-)reciprocity of the system, we can make the following observations. Increasing the total density ($\rho_0^A+\rho_0^B$) generally favors flocking. On the other hand, the composition ($\rho_0^A-\rho_0^B$) affects the relative alignment strength of particles. This can be seen from the continuum Eqs.~(\ref{eq:continuum_eq_density})-(\ref{eq:continuum_eq_polarization}), where the alignment parameter $g'_{ab}\sim g_{ab}\,\rho_0^b$. Thus, even for constant $g_{ab}$, the total torque depends on $\rho_0^A-\rho_0^B$.

The phase diagram for a reciprocal system is shown in Fig.~\ref{fig:impact_parameter_rho_reciprocal}. Here, $g_{AA}=g_{BB}=3$ and $g_{AB}=g_{BA}=6$. Symmetric clustering only emerges in equal-density mixtures with $\rho_0^A-\rho_0^B=0$. For $\rho_0^A \neq \rho^B_0$, asymmetric clustering is predicted -- even in the reciprocal limit.

The phase diagram for a non-reciprocal system with $g_{AA}=g_{BB}=3$ and $g_{AB}=-g_{BA}=3$ is shown in Fig.~\ref{fig:impact_parameter_rho_non-reciprocal}. For equal-density mixtures, asymmetric $A$-clustering is predicted. Nevertheless, for $\rho_0^A\ll\rho^B_0$, even asymmetric $B$-clustering can occur.

To reduce complexity in our model, we only consider the simplest case of an equal-density mixture in the main text.

\section{\label{app:correlation_functions}Pair correlation functions}
Information on the translational structure in our active binary mixture are captured by the pair correlation function $G_{ab}(\bm{r})$, which describes the distribution of distance vectors $\bm{r}$ between pairs of particles belonging to species $a$ and $b$ \cite{Bartnick_Loewen_2015_structural_correlations_in_nonreciprocal_systems}. In homogeneous systems, we define $G_{ab}(\bm{r})$ as \cite{hansen_mcDonald_2013_theory_simple_liquids,Bartnick_Loewen_2015_structural_correlations_in_nonreciprocal_systems}
\begin{equation}
	\label{eq:pair_distribution_function}
	G_{ab}(\bm{r}) = \frac{1}{\Omega} \, \sum_{a_i=1}^{N_{a}} \sum_{\substack{b_j=1\\ (b_j\neq a_i)}}^{N_{b}} \Big \langle \delta(\bm{r} - (\bm{r}_{a_i} - \bm{r}_{b_j})) \Big \rangle,
\end{equation}
where $\Omega = N_a\,N_b/V$ is the normalization and $V=L^2$ represents the volume of the system. $G_{ab}(\bm{r})$ tends to unity for $r\rightarrow \infty$ and vanishes for $r \rightarrow 0$ due to steric repulsion between particles. 

Numerically, we determine $G_{ab}(r,\phi)$ by counting the particles found in small area fractions of distance $r+\Delta r$ and angle $\phi + \Delta \phi$ from the reference particle, such that we additionally normalize with the area fraction element $\Delta A = r\,\Delta r \, \Delta\phi$, leading to 
\begin{equation}
	G_{ab}(r,\phi) = \frac{1}{\Omega_{\rm n}} \, \sum_{a_i=1}^{N_{a}} \sum_{\substack{b_j=1\\ (b_j\neq a_i)}}^{N_{b}} \Big \langle \delta(r^{ab}_{ij} - r)\,\delta({\phi^{ab}_{ij}-\phi}) \Big \rangle
\end{equation}
with $\Omega_{\rm n} = N_{a}\,N_{b}\,\Delta A/V$. The relative particle position and angle are calculated as $r^{ab}_{ij} = \vert \bm{r}_{b_j} - \bm{r}_{a_i} \vert$ and \mbox{$\phi_{ij}^{ab} = \sphericalangle(\bm{r}_{b_j}-\bm{r}_{a_i}, \bm{p}_{a_i})$}.

Note that by construction, the pair distribution function is symmetric in the sense that $G_{AB} = G_{BA}$. The effect of non-reciprocity is expected to manifest itself in differences between the single-species correlations, such that $G_{AA} \neq G_{BB}$ \cite{Bartnick_Loewen_2015_structural_correlations_in_nonreciprocal_systems}. We here focus on the radial correlations $G_{ab}(r) = \langle G_{ab}(r,\phi) \rangle_{\phi}$, given as the average over all relative angles $\phi$. 
Details and further examples of pair correlation functions are shown in \cite{kreienkamp_klapp_PRE}.

\section{Structure factor matrix}
To characterize the density fluctuations close to phase transitions in our binary mixture, we took inspiration from established procedures applied in equilibrium mixtures \cite{chen_forstmann_1992_demixing_binary_Yukawa_fluid, chen_forstmann_1992_phase_instability_fluid_mixtures}. Our approach involves the computation of density fluctuation correlations of form $\langle \delta \rho_{a}(\bm{r}) \, \delta \rho_{b}(\bm{r}') \rangle$, where $a,b = A,B$. Here, we only consider instantaneous fluctuations and neglect all time-dependencies. The density fluctuation is given by $\delta \rho_{a}(\bm{r}) = \rho_{a}(\bm{r}) - \rho_{0}^a$ with $\rho_{0}^a$ as the density of the homogeneous system. In an additionally isotropic system, the density fluctuations in Fourier space (denoted by a hat, $\hat{\cdot}$ \footnote{We define the (two-dimensional) Fourier transform as $\hat{f}(\bm{k}) =  \int_{\mathbb{R}^2} f(\bm{r}) \, {\rm e}^{-2\pi i\bm{r}\cdot \bm{k}} {\rm d}\bm{r}$ (with inverse $f(\bm{r}) =  \int_{\mathbb{R}^2} \hat{f}(\bm{k}) \, {\rm e}^{2\pi i\bm{r}\cdot \bm{k}} {\rm d}\bm{k}$). For a radially symmetric integral kernel, the two-dimensional Fourier transform is a Hankel or Fourier-Bessel transform (of order zero) with $\hat{f}(k) = 2\,\pi \int_{0}^{\infty} f(r) \, J_0(2\,\pi\,k\,r)\,r\,{\rm d}r$, where $J_0(z)$ is the zeroth order Bessel function of the first kind.}) read \cite{kreienkamp_klapp_PRE}
\begin{equation}
	\label{eq:mean_fluctuation_products}
	\frac{1}{V}\,\langle \delta \hat{\rho}_{a}(\bm{k}) \, \delta \hat{\rho}_{b}(-\bm{k}) \rangle = \rho_{0}^a \,\rho_{0}^b\, \hat{h}_{ab}(k) + \delta_{ab} \, \rho_{0}^a,
\end{equation}
where $h_{ab}(\vert \bm{r}-\bm{r}'\vert)=G_{ab}(\vert \bm{r}-\bm{r}'\vert) - 1$ denotes the total correlation function \cite{hansen_mcDonald_2013_theory_simple_liquids}.
We note already here that in the present system, the assumption of homogeneity and isotropy holds only for short times (after starting from a random configuration).

To characterize the type of phase transition within the binary mixture, we consider two different types of fluctuations: fluctuations in the total density $\delta\hat{\rho}(k) = \delta \hat{\rho}_A(k) + \delta \hat{\rho}_B(k)$ and fluctuations in the composition $\delta\hat{c}(k) = \delta \hat{\rho}_A(k) - \delta \hat{\rho}_B(k)$. These fluctuations can be written in terms of the structure factor matrix $\bm{\mathcal{S}}(k)$, given by
\begin{equation}
	\label{eq:structure_factor_matrix}
	\bm{\mathcal{S}}(k) = \begin{pmatrix}
		S_{\rho\rho}(k) & S_{c\rho}(k) \\
		S_{c\rho}(k) & S_{cc}(k)
	\end{pmatrix}
\end{equation}
with matrix elements
\begin{equation}
	\begin{split}
		S_{\rho\rho}(k) &= \frac{1}{V}\,\langle \delta\hat{\rho}(\bm{k}) \, \delta\hat{\rho}(-\bm{k}) \rangle\\
		& = (\rho_{0}^A)^2 \, \hat{h}_{AA}(k) + (\rho_{0}^B)^2 \, \hat{h}_{BB}(k) + \rho_{0}^A + \rho_{0}^B + 2\,\rho_{0}^A\,\rho_{0}^B\,\hat{h}_{AB}(k),
	\end{split}
\end{equation}
\begin{equation}
	\begin{split}
		S_{cc}(k) &= \frac{1}{V}\,\langle \delta\hat{c}(\bm{k}) \, \delta\hat{c}(-\bm{k}) \rangle\\
		& = (\rho_{0}^A)^2 \, \hat{h}_{AA}(k) + (\rho_{0}^B)^2 \, \hat{h}_{BB}(k)   + \rho_{0}^A + \rho_{0}^B - 2\,\rho_{0}^A\,\rho_{0}^B\,\hat{h}_{AB}(k),
	\end{split}
\end{equation}
and
\begin{equation}
	\begin{split}
		S_{c\rho}(k) &= S_{\rho c}(k) = \frac{1}{V}\, \langle \delta\hat{c}(\bm{k}) \, \delta\hat{\rho}(-\bm{k}) \rangle\\
		& = (\rho_{0}^A)^2 \, \hat{h}_{AA}(k) - (\rho_{0}^B)^2 \, \hat{h}_{BB}(k)  + \rho_{0}^A - \rho_{0}^B .
	\end{split}
\end{equation}
We assume that, as in equilibrium, an instability related to a phase transition is signaled by the divergence of fluctuations. This means, that one eigenvalue $\lambda_{1/2}(k)$ of $\bm{\mathcal{S}}(k)$ diverges at the transition. In particular, symmetric clustering (condensation) is characterized by diverging fluctuations in the total density. A demixing phase transition corresponds to diverging fluctuations in the composition. Consequently, the eigenvalues $\lambda_{1/2}(k)$ and corresponding (normalized) eigenvectors $\bm{v}_{1/2}(k)=(\delta \hat{\rho}(k),\, \delta \hat{c}(k))^{\rm T}$ of matrix $\bm{\mathcal{S}}(k)$ indicate whether and what type of phase transition occurs. More specifically, when $\lambda_1^{-1}(k)$ or $\lambda_2^{-1}(k)$ goes to zero, the respective eigenvector $\bm{v}_{\rm max}$ indicates whether the phase transition is predominantly symmetric clustering ($\bm{v}_{\rm max}\approx \bm{x}_{\rm con} = (1,0)^{\rm T}$) or de-mixing ($\bm{v}_{\rm max}\approx \bm{x}_{\rm dem} =(0,1)^{\rm T}$). We quantify the degree of symmetric clustering and/or demixing in terms of the angle $\alpha = \arccos(\bm{v}_{\rm max} \cdot \bm{x}_{\rm con})$ between the eigenvector $\bm{v}_{\rm max}$ and the vector $\bm{x}_{\rm con}$, representing symmetric clustering.

Besides symmetric clustering ($\alpha=0$) and demixing ($\alpha=\pi/2$), the angle $\alpha$ also indicates whether rather species $A$ or $B$ forms clusters. In particular, $0<\alpha<\pi/2$ corresponds to asymmetric clustering of species $A$ and $-\pi/2 < \alpha < 0$ corresponds to asymmetric clustering of species $B$.

In our analysis, it turns out that the limit $k\rightarrow 0$ is the most relevant since $\lambda_{1/2}^{-1}$ are smallest there. Therefore, the presented results refer exclusively to this limit.

\section{List of supplementary videos}
\label{sec:list_videos}
To visualize the dynamics of non-equilibrium phases exhibited in the binary mixture, we present videos of our BD simulations. They represent one exemplary non-equilibrium steady state for a single random initial configuration, respectively. The videos show
\begin{itemize}
	\item the (reciprocal) flocking state for $g_{AB}=g_{BA}=9$,
	\item the (reciprocal) anti-flocking state for $g_{AB}=g_{BA}=-9$,
	\item the (reciprocal) symmetric clustering state for $g_{AB}=g_{BA}=1$,
	\item the (non-reciprocal) asymmetric $B$-clustering state for $g_{AB}=-g_{BA}=-9$,
	\item the (non-reciprocal) asymmetric $A$-clustering state for $g_{AB}=6$, $g_{BA}=-9$,
	\item the (non-reciprocal) disordered for $g_{AB}=-g_{BA}=25$.
\end{itemize}
The intraspecies couplings are set to $g_{AA}=g_{BB}=3$. Other parameters are chosen as described in the Sec.~\ref{app:microscopic_model}.

\input{supplemental_material.bbl}

%% file: main.bbl
%

%% file: supplemental_material.bbl
%